\documentclass[%
 aip,
 jmp,%
 amsmath,amssymb,
 reprint,%
]{revtex4-1}

\usepackage{graphicx}
\usepackage{dcolumn}
\usepackage{bm}
\usepackage{subfigure}
\usepackage[english]{babel}
\usepackage{lipsum}

\usepackage{soul}
\usepackage{amsmath,amssymb,amsfonts}
\usepackage{graphics}
\usepackage{color}
\usepackage{xcolor}
\definecolor{b}{rgb}{0,0,0}
\definecolor{rev}{rgb}{0,0,0}

\usepackage{amsthm}
\theoremstyle{plain}

\theoremstyle{definition}

\theoremstyle{remark}
\newtheorem*{remark}{Remark}

\usepackage{mathtools, nccmath}
\usepackage{algorithm}
\usepackage{algorithmic}
\usepackage{enumitem}
\setlist[itemize]{leftmargin=*}

\usepackage{comment}
\usepackage{hyperref}
\usepackage{lipsum}
\usepackage{tabularx}
\usepackage{mathrsfs}
\usepackage{stackengine}

\newcommand{\ba}{\boldsymbol{a}}
\newcommand{\bx}{\boldsymbol{x}}
\newcommand{\bff}{\boldsymbol{f}}

\begin{document}



\title{On closures for reduced order models -- A spectrum of first-principle to machine-learned avenues}


\author{Shady E. Ahmed}
\author{Suraj Pawar}
\author{Omer San}
\email{osan@okstate.edu}
\affiliation{ 
School of Mechanical \& Aerospace Engineering, Oklahoma State University, Stillwater, OK 74078, USA.
}

\author{Adil Rasheed}
\affiliation{ 
Department of Engineering Cybernetics, Norwegian University of Science and Technology, N-7465, Trondheim, Norway.
}
 
\author{Traian Iliescu}
\affiliation{ 
Department of Mathematics, Virginia Tech, Blacksburg, VA 24061, USA.
}

\author{Bernd R. Noack}
\affiliation{School of Mechanical Engineering and Automation, Harbin Institute of Technology, Shenzhen 518058, China.}%
\affiliation{Hermann-F\"ottinger-Institut f\"ur Str\"omungsmechanik, Technische Universit\"at Berlin, D-10623 Berlin, Germany}

\date{\today}

\begin{abstract}

For over a century, reduced order models (ROMs) have been a fundamental discipline of theoretical fluid mechanics. Early examples include Galerkin models inspired by the Orr-Sommerfeld stability equation and numerous vortex models, of which the von K\'arm\'an vortex street is one of the most prominent. Subsequent ROMs typically relied on first principles, like mathematical Galerkin models, weakly nonlinear stability theory, and two- and three-dimensional vortex models. Aubry et al. [N. Aubry, P. Holmes, J. Lumley,  and E. Stone, Journal of Fluid Mechanics, 192, 115–173 (1988)] pioneered data-driven proper orthogonal decomposition (POD) modeling. In early POD modeling, available data was used to build an optimal basis, which was then utilized in a classical Galerkin procedure to construct the ROM. But data has made a profound impact on ROMs beyond the Galerkin expansion. In this paper, we take a modest step and illustrate the impact of data-driven modeling on one significant ROM area. Specifically, we focus on ROM closures, which are correction terms that are added to classical ROMs in order to model the effect of the discarded ROM modes in under-resolved simulations. Through simple examples, we illustrate the main modeling principles used to construct classical ROMs, motivate and introduce modern ROM closures, and show how data-driven modeling, artificial intelligence, and machine learning have changed the standard ROM methodology over the last two decades. Finally, we outline our vision on how state-of-the-art data-driven modeling can continue to reshape the field of reduced order modeling.

\end{abstract}


\maketitle

\section{Introduction} \label{sec:intro}
One of the very first exciting experiences that kids go through is playing with water; they might throw a stone in a lake, float a rubber duck in a bathtub, or even stir a straw while enjoying a tasty cup of juice! They like doing this over and over again because of the magnificent patterns that keep forming every time. These patterns or \emph{coherent structures} are ubiquitous in the world, in general, and in fluid flows, in particular. They attracted Leonardo da Vinci more than five centuries ago, resulting in some of his outstanding artwork \cite{marusic2021leonardo}. Fluid dynamicists are especially lucky to enjoy the beauty of these formations on a daily basis. But other than their aesthetic value, these patterns come with a practical benefit. In particular, these coherent structures are the cornerstone in the development of {\it reduced order models (ROMs)} for fluid flows. ROMs are built by using available {\it data} to identify and rank these structures, choosing the most effective few of them, and tracking their dynamical behavior in order to approximate the evolution of the underlying flow. The computational cost of the relatively low-dimensional ROMs is dramatically lower than the computational cost of a direct numerical simulation, which aims at capturing all the flow scales. Since their introduction to the field of fluid dynamics more than fifty years ago~\cite{lumley1967structure}, ROMs have witnessed tremendous changes. Arguably, {\it data-driven modeling} has been the main driving force behind these changes. Over the last two decades, state-of-the-art methods from {\it machine learning (ML)} have reshaped the field of reduced order modeling.

The main objective of this study is to provide an overview of data-driven reduced order modeling strategies relevant to fluid dynamics applications. The topic spans a wide spectrum, and there are many review articles on pertinent discussions, methodologies, and applications in fluids \cite{kou2021data,beck2021perspective,brunton2020machine,taira2020modal,mendoncca2019model,yu2019non,brenner2019perspective,yondo2019review,taira2017modal,rowley2017model,brunton2015closed,lassila2014model,mezic2013analysis,chinesta2011short,massarotti2010reduced,lucia2004reduced,berkooz1993proper} as well as closely related fields
\cite{san2021hybrid,heinlein2021combining,blechschmidt2021three,kashinath2021physics,bauer2021digital,frank2020machine,zhu2020river,tahmasebi2020machine,benner2015survey,asher2015review,machairas2014algorithms,mignolet2013review,bazaz2012review,razavi2012review,theodoropoulos2011optimisation,wagner2010model,kleijnen2009kriging,pinnau2008model,ong2003evolutionary,freund2003model,bai2002krylov,chatterjee2000introduction,bonvin1982unified,elrazaz1981review}. Therefore, it is not our intention to include a detailed discussion, but rather to survey one important ROM research area, closure modeling, and provide our {\it subjective} perspectives on how data-driven modeling has made an impact in this area. In particular, given the recent interest in ML applications in fluid dynamics, our survey is intended to encourage cross-disciplinary efforts between practitioners, physicists, mathematicians, and data scientists. We hope that our paper will shed light on the new ideas of integrating both physics in ML models and ML-enabled capabilities in principled models, a rapidly emerging field that came to be known as  physics-guided ML (PGML). These developments are born to conform with the scientific foundations that are moving rapidly to the industry to enable the next generation of digital twin technologies \cite{rasheed2020digital}. 

\begin{figure*}[ht]
\centering
\includegraphics[width=1.0\linewidth]{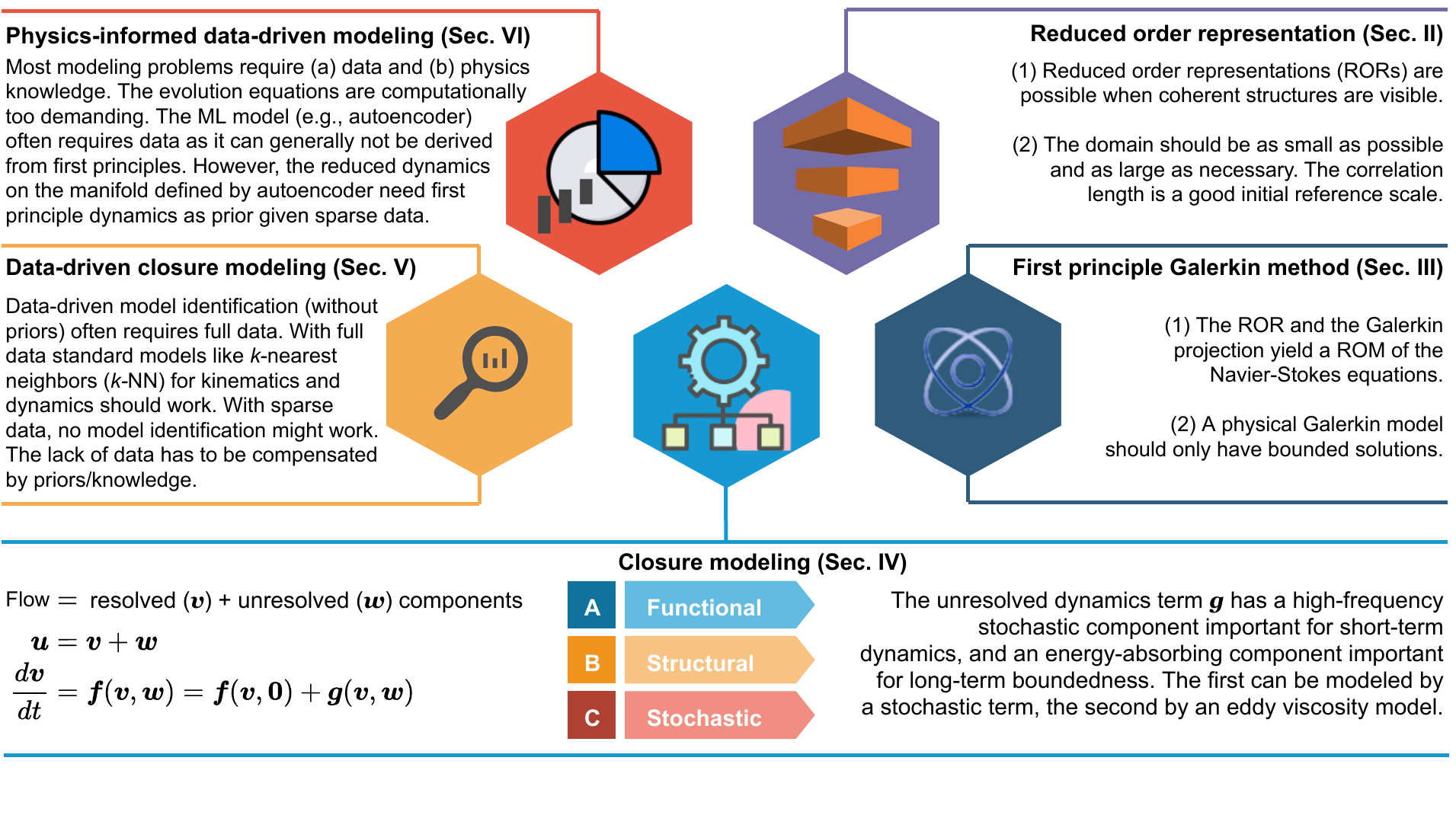} \vspace{-20pt}
\caption{An overview of data-driven reduced order modeling in fluid dynamics.}
\label{fig:pgai}
\end{figure*}

This paper first aims at identifying imminent practical and mathematical needs in designing closure approaches for ROMs of nonlinear parameterized fluid dynamics systems, i.e., complex natural or engineered systems comprising coupled partial differential equations (PDEs) with variable parameters, and initial and boundary conditions \cite{karatzas2020projection}. Such models usually serve as the inner-workhorse for outer-workflow loops such as optimal design \cite{aguado2017simulation}, control \cite{aguado2015real,zainib2019reduced}, estimation, and discovery \cite{quarteroni2014reduced,benner2015a,wang2020deep,swischuk2019projection,lee2020model}. In particular, there has recently been an increasing interest in ROMs from the fluid dynamics community, where emerging data-driven methods prevail. This is primarily due to the fact that data and centralized powerful open-source machine learning and optimization libraries have become widespread, as indicated in Figure~\ref{fig:pgai}. Although we mainly focus on incompressible flows, we emphasize that there have been inspiring works done in the compressible case \cite{iollo2000stability,rowley2004model,bourguet2007reduced,gloerfelt2008compressible,bourguet2009capturing,barone2009stable,kalashnikova2010stability,placzek2011nonlinear,balajewicz2016minimal,ferrero2018global,yano2019discontinuous,yu2019data,rezaian2020hybrid,zucatti2021calibration,krath2021efficient}. 

To begin with, reduced order modeling can be viewed as the art of converting existing prior information and collected data into a dramatically more efficient, yet relatively accurate, surrogate model to be used on demand. For example, a conceivable strategy of flow control is to put most of the demanding calculations offline and to keep only low rank updates for fluid flow evolutions online \cite{hovland2008explicit,ballarin2020reduced}. Emerging digital twin infrastructures are one of the main beneficiaries and driving forces behind efficient surrogate model development efforts \cite{kapteyn2020data,kapteyn2021probabilistic}. Although the ROM concept is not new in fluid dynamics, there are still many new fronts and opportunities, mainly due to the recent advances in ML algorithms and easy-to-use open-source packages that can be utilized in many control and optimization processes. We also note that, in many fluid dynamics applications, the typical data sparsity (due to the number of necessarily resolved degrees of freedom being orders of magnitude larger than the available sensors) and corruption (e.g., due to signal noise, interference, and sensor malfunctioning) motivate physics informed data-driven modeling.

Among fluid dynamicists, projection-based linear methods have become popular. Both proper orthogonal decomposition (POD) \cite{holmes1998turbulence} and dynamic mode decomposition (DMD)\cite{kutz2016dynamic} enabled approaches have been exploited. In our work, we mostly focus on POD relevant literature and refer the reader to \citet{kutz2016dynamic} for the DMD principles. The mathematical foundations behind the POD-based linear subspace approaches go back to the principal component analysis (PCA), pioneered by Karl Pierson \cite{pearson1901principal} in 1901 and later demonstrated graphically by Harold Hotelling \cite{hotelling1933analysis} in 1933. This powerful statistical approach (also known as Kosambi-Karhunen-Lo\`{e}ve expansion\cite{kosambi2016statistics,loeve1955probability} or empirical orthogonal functions\cite{lorenz1956empirical,monahan2009empirical}) was first introduced in the fluid dynamics community by John Lumley \cite{lumley1967structure,lumley1972stochastic,lumley1981coherent}, and came to be  known as POD. In practice, the \emph{method of snapshots}, established by Lawrence Sirovich \cite{sirovich1987turbulence}, was a key enabler to efficiently determine the POD modes for large scale problems, often encountered in fluid dynamics. Of particular interest when characterizing the dynamics of coherent structures in wall bounded flows, the beauty of the POD modeling approach was demonstrated in seminal works by Nadine Aubry \cite{aubry1988dynamics,aubry1991hidden}. An admittedly incomplete chronological evolution of projection-based ROMs is given in Table~\ref{table:keypapers}. 

A key advantage of POD is the guaranteed minimal representation error for the employed snapshots with respect to all other Galerkin expansions with the same number of modes. Another advantage is the orthogonality of the constructed modes that naturally leverages the use of a Galerkin type projection onto the governing PDEs to obtain a system of ordinary differential equations, defining a dynamical system for the amplitudes of POD modes. These models are often \emph{intrusive} in the sense that both the governing equations and the prerecorded snapshot data are required to build a ROM. Postulating an alternative approach, DMD based approaches and many other \emph{nonintrusive} models bypass this equation dependency in order to construct the ROM solely based on the prerecorded snapshots \cite{willcox2002balanced,barrault2004empirical,bai2002krylov,mezic2005spectral,grepl2007efficient,gugercin2008h2,chaturantabut2010nonlinear,schmid2010dynamic,williams2015data,carlberg2011efficient,tu2014dynamic,lassila2014model,benner2015survey,brunton2015closed}. 

\begin{table*}[htbp]
    \centering
    \caption{An incomplete chronological list of key contributions to projection-based ROMs in fluid dynamics.}
    \begin{tabular}{p{0.05\linewidth}p{0.34\linewidth}p{0.58\linewidth}}
    \hline\noalign{\smallskip}
    Year & Study & Key Contribution \\
    \noalign{\smallskip}\hline\noalign{\smallskip}
    1915 & \citet{galerkin1915series} & Galerkin method for solving (initial) boundary value problems \\
    1962 & \citet{saltzman1962finite} & Low-dimensional modeling (with 7 modes, see also Ref\cite{lakshmivarahan2019saltzman} for a revisit) \\
    1963 & \citet{lorenz1963deterministic} & Low-dimensional modeling (with 3 modes) \\
    1967 & \citet{lumley1967structure} & Proper orthogonal decomposition (POD) \\ 
    1987 & \citet{sirovich1987turbulence} & Method of snapshots \\
    1988 & \citet{aubry1988dynamics} & First POD model:  Dynamics of coherent structures and global eddy viscosity modeling\\
    1994 & \citet{rempfer1994dynamics} & Linear modal eddy viscosity closure\\
    1995 & \citet{everson1995karhunen} & Gappy POD \\
    2000 & \citet{ravindran2000reduced} & Galerkin ROM for optimal flow control problems \\
    2001 & \citet{kunisch2001galerkin} & First numerical analysis of Galerkin ROM for parabolic problems \\
    2002 & \citet{willcox2002balanced} & Balanced truncation with POD \\
    2003 & \citet{couplet2003intermodal}  & Guidelines for modeling unresolved modes in POD–Galerkin models \\
    2004 & \citet{sirisup2004spectral} & Spectral viscosity closure for POD models \\
    2004 & \citet{barrault2004empirical} & Empirical interpolation method (EIM) \\
    2005 & \citet{mezic2005spectral} & Spectral decomposition of the Koopman operator\\
    2007 & \citet{rozza2007reduced} & Reduced basis approximation \\
    2007 & \citet{cao2007reduced} & Galerkin ROM for four‐dimensional variational data assimilation \\
    2008 & \citet{amsallem2008interpolation} & Interpolation method based on the Grassmann manifold approach \\
    2008 & \citet{astrid2008missing} & Missing point estimation\\
    2009 & \citet{rowley2009spectral} & Spectral analysis of nonlinear flows \\
    2009 & \citet{sapsis2009dynamically} & \textcolor{rev}{Dynamically orthogonal field equations} \\
    2010 & \citet{schmid2010dynamic} & A purely nonintrusive perspective: Dynamic mode decomposition (DMD) \\
    2010 & \citet{chaturantabut2010nonlinear} & Discrete empirical interpolation method (DEIM)\\
    2013 & \citet{carlberg2013gnat} & The Gauss–Newton with approximated tensors (GNAT) method \\
    2013 & \citet{cordier2013identification} &  Proof of global boundedness of nonlinear eddy viscosity closures\\
    2014 & \citet{osth2014need} & $\sqrt{K}$-scaled eddy viscosity concept\\
    2015 & \citet{ballarin2015supremizer} & Stabilization of POD Galerkin approximations\\
    2015 & \citet{schlegel2015long} & On bounded solutions of Galerkin models \\
    2016 & \citet{peherstorfer2016data} & Data-driven operator inference nonintrusive ROMs \\
    2016 & \citet{brunton2016discovering} & \textcolor{rev}{Sparse identification of nonlinear dynamics (SINDy)} \\
    2016 & \citet{sieber2016spectral} & Spectral POD \\
    2018 & \citet{towne2018spectral} & On the relationship between spectral POD, DMD, and resolvent analysis \\
    2018 & \citet{reiss2018shifted} & Shifted/transported snapshot POD \\
    2018 & \citet{loiseau2018sparse} & Feature-based manifold modeling \\
    2019 & \citet{mendez2019multi} & Multi-scale proper orthogonal decomposition\\
    2021 & \citet{li2021cluster}, \citet{fernex2021cluster} & Cluster-based network models \\
    \noalign{\smallskip}\hline
    \end{tabular}
    \label{table:keypapers}
\end{table*}

If we adopt the training and testing terminologies from ML, both POD and DMD based models require the \emph{training} snapshot data to build the ROM (i.e., data-driven modeling). The fundamental question in practice is how well these ROMs will perform in testing conditions (i.e., the conditions that are not included in the training data). The trade-off performance between training and testing constitutes one of the crucial questions about the credibility of the proposed ROMs, and motivates more efforts, ideas, and collaborations to push the frontiers of existing ROM frameworks. Physics informed ML has also made an impact in reduced order modeling. In this hybrid approach, while the training data provides a set of global basis functions, the underlying governing equations (i.e., physics) constrain the evolution within the linear subspace defined by these POD basis functions. Deep discussions on hybrid approaches that combine deterministic and statistical modeling can be found elsewhere \cite{krasnopolsky2006new,krasnopolsky2006complex,reichstein2019deep,de2019deep,karpatne2017theory,childs2019embedding,willard2020integrating,san2021hybrid}. 

Projection-based ROMs have been explored for decades, and these explorations have been paying off in many applications. They have great promise for flow control of industrial processes, for enlarging ensemble size for flow problems with uncertain data, and even for providing accurate forecasts of fluid behaviour. The apparent success of low-rank ensemble nonlinear filtering methods utilized in weather forecasting centers also suggests that there is a prospect of using a system whose dimension is substantially lower than the dimension of the state space. Yet, the ROMs' potential has been only realized for a small collection of canonical flows. One of the main roadblocks for ROMs of realistic flows is that they are not accurate models for the dominant modes. In practice, a closure or correction term is generally added \cite{kalb2007intrinsic,amsallem2012stabilization,balajewicz2012stabilization,wang2012proper,cordier2013identification,osth2014need,protas2015optimal,benosman2017learning,stabile2019reduced,reyes2020projection}. In many cases, there are complementary physical, statistical, and computational challenges that arise in the development of ROMs and ROM closures, topics that we will systematically survey in this work toward establishing foundations to close the gap between what ROMs can do and where they are needed.

\section{Reduced Order Representation} \label{sec:ROR}

A \emph{reduced order representation} (ROR) can be viewed as a generalization of the latent space or manifold,  i.e., a simplifying kinematic approximation. For example, the POD procedure introduced in Section~\ref{sec:POD} constitutes a best-fit linear manifold to establish a ROR. The ROR facilitates a data compression for an ensemble of snapshot data. Physically interpretable RORs are possible when dominant coherent structure are present. \textcolor{rev}{This is clearly ubiquitous in the fluid flows that we encounter in our daily life, as introduced in Section~\ref{sec:intro}, as well as large-scale and industrial settings. For spatio-temporal dynamical systems, the rank-r ROR of the state $u(\boldsymbol x,t)$ can be simply written as
\begin{equation}
u(\boldsymbol x,t) = \sum_{i=1}^r a_i(t) \psi_i(\boldsymbol{x}), \label{eq:ror}
\end{equation}
where $\boldsymbol x$ refers to the spatial coordinates, $t$ is the time, $\psi_i$ denotes the $i$-th mode in the ROR, and $a_i$ is the corresponding amplitude or coefficient. Although it is generally assumed that the basis functions $\psi_i$ are time-independent and the dynamical evolution is encapsulated in the coefficients $a_i$, there have been studies that admit a time-evolving basis functions as well \cite{sapsis2009dynamically,cheng2013dynamically,ramezanian2021fly,patil2020real}. 
}

\textcolor{rev}{\subsection{Eigenfunction expansion} \label{sec:eig}
Any set of $n$ linearly independent vectors can serve as a basis for an $n$-dimensional vector space. Any vector in this space can be expressed as a linear combination of these linearly independent basis vectors. In an infinite dimensional vector space of functions, there exists an infinite set of linearly independent basis functions $\{\psi_i(\boldsymbol x)\}_{i=1, 2, \dots}$ such that a given function $u(\boldsymbol x)$ in this space can be written as a linear combination of these functions. It is straightforward to show that any periodic, piecewise continuous function can be written as an infinite sum of sines and cosines (e.g., Fourier series \cite{haberman2012applied}). The eigenfunction expansion can be viewed as a generalization of the Fourier series expansion for arbitrary boundary conditions, where the Sturm-Liouville theory provides an infinite sequence of eigenvalue-eigenfunction pairs. 
ROR also seeks an expansion of an arbitrary function in terms of a given set of basis functions.  However, in contrast to the methods mentioned in this section, ROR aims at finding a low-dimensional basis instead of an infinite dimensional one.
}

\textcolor{b}{To illustrate these concepts, let us consider a linear advection problem in a spatial domain $[0, L]$:
\begin{equation} \label{eq:wave}
\frac{\partial u}{\partial t} + c\frac{\partial u}{\partial x} = 0,
\end{equation}
where $c$ is the wave speed. To select the appropriate basis functions, we consider the boundary conditions. For example, if we have homogeneous Dirichlet boundary conditions, i.e.,
\begin{equation} \label{eq:wave2}
u(x=0,t) = 0, \quad u(x=L,t) = 0, \quad \mbox{for} \quad t \in [0,T],
\end{equation}
we might choose a set of \emph{orthonormal} basis functions $\psi_i(x)$ defined as 
\begin{equation} \label{eq:wave3}
\psi_i(x) = \sqrt{\frac{2}{L}}\sin\Big(\frac{i\pi}{L}x\Big),
\end{equation}
in order to approximate $u$ as follows:
\begin{equation} \label{eq:sinepsi}
u(x,t) = \sum_{i=1}^{r}a_{i}(t) \psi_i(x).
\end{equation}
}

\textcolor{b}{In a more sophisticated scenario with the homogeneous Neumann condition on the left boundary ($x=0)$ and the homogeneous Drichlet condition on the right boundary ($x=L=1$), i.e.,
\begin{equation} \label{eq:wave4}
\frac{\partial u}{\partial x} \Big|_{x=0} = 0, \quad u\Big|_{x=1} = 0, \quad \mbox{for} \quad t \in [0,T],
\end{equation}
we can define a set of \emph{non-orthogonal} functions that satisfy Eq.~\ref{eq:wave4} as follows:
\begin{equation} \label{eq:wave5}
\phi_i(x) = \cos(i\pi x) - (-1)^{i}
\end{equation}
and apply the Gram-Schmidt \emph{orthonormalization} process to obtain the following set of basis functions:
\begin{eqnarray} \label{eq:wave6}
\psi_i(x) &= \sqrt{\frac{4i-2}{2i+1}}\Bigg[\frac{(-1)^{i+1}}{2i-1} +\cos(i \pi x) \nonumber \\ & + \frac{2}{2i-1} \sum_{j=1}^{i-1} (-1)^{i+j+1}\cos(j \pi x)\Bigg].
\end{eqnarray}
Here, we note that these basis functions are orthonormal, i.e.,
\begin{equation} \label{eq:ort}
\int_{0}^{1} \psi_i(x) \psi_k(x)dx = \delta_{ik} , 
\end{equation}
and they are derived from the Fourier harmonics that satisfy the boundary conditions.}

\textcolor{b}{
Next, we focus on an illustrative example with periodic boundary conditions, with the domain length $L = 2 \pi$, maximum time $T=2 \pi$, and wave speed $c=1$. For a given initial condition $u(x,t=0) = \cos(x)$, Eq.~\ref{eq:wave} admits an analytical solution in the form of a right travelling wave $u(x,t) = \cos(x-t)$. Let us approximate $u$ using a modal expansion with only two Fourier harmonics defined by 
\begin{equation} \label{eq:wave7}
u(x,t) = a_{1}(t)\cos(x) + a_{2}(t)\sin(x).
\end{equation}
}
\textcolor{b}{
Substituting Eq.~\ref{eq:wave7} into Eq.~\ref{eq:wave}, we get
\begin{eqnarray} \label{eq:wave8}
& \frac{\partial}{\partial t} \Big(a_{1}(t)\cos(x) + a_{2}(t)\sin(x)\Big)  \nonumber \\ & + \frac{\partial}{\partial x} \Big(  a_{1}(t)\cos(x) + a_{2}(t)\sin(x)\Big) = 0.
\end{eqnarray}
Once we multiply Eq.~\ref{eq:wave8} with $\cos(x)$ and integrate over the domain, we obtain an equation for $a_{1}(t)$,
\begin{eqnarray} \label{eq:wave9}
\frac{d a_{1}}{d t} = - a_{2},
\end{eqnarray}
and similarly, multiplying Eq.~\ref{eq:wave8} with $\sin(x)$, the evolution equation for $a_2$ becomes 
\begin{eqnarray} \label{eq:wave10}
\frac{d a_{2}}{d t} =  a_{1}.
\end{eqnarray}
Eq.~\ref{eq:wave9} and Eq.~\ref{eq:wave10} constitute the well-known Galerkin system. Using the initial condition given at $t=0$,
\begin{equation} \label{eq:wave11}
a_{1}(0) = 1, \quad a_{2}(0) = 0,
\end{equation}
we can obtain an analytical solution of the Galerkin system given by Eq.~\ref{eq:wave9} and Eq.~\ref{eq:wave10} as
\begin{equation} \label{eq:wave12}
a_{1}(t) = \cos(t), \quad a_{2}(t)=\sin(t).
\end{equation}
Therefore, the two-mode Galerkin model approximation given by Eq.~\ref{eq:wave7} yields a solution
\begin{equation} \label{eq:wave13}
u(x,t) = \cos(t)\cos(x) + \sin(t)\sin(x),
\end{equation}
which can be further written as 
\begin{equation} \label{eq:wave14}
u(x,t) = \cos(x-t).
\end{equation}
As we illustrated in this example, the two-mode approximation retrieves the exact solution. One of the key aspects in such a modal scheme is therefore related to the characteristics of the selected basis functions, which ultimately provided the best possible expansion in this example. A central question is how we would know a priori the appropriate $\cos(x)$ and $\sin(x)$ basis functions to approximate $u$.}

\textcolor{rev}{We also note that the multimodal method utilizes similar arguments to represent the solution as a superposition of an infinite set of generalized Fourier bases and time-dependent coefficients. The multimodal method has been extensively exploited to study the sloshing problem \cite{faltinsen2009sloshing,lukovsky2017multimodal,gavrilyuk2000multimodal}, where a set of natural harmonic functions are defined to satisfy the boundary conditions. This definition is challenging since each individual tank shape requires a dedicated applied mathematical and physical study. Moreover, \citet{faltinsen2001adaptive} reported that the simple limitation of the infinite sum by a finite number $r$ can yield either inaccurate or expensive computations. Thus, the selection of the dominant modes is a non-trivial task. A truncation based on employing special asymptotic relationships, postulated following  mathematical or physical arguments, has been shown to produce good results for the sloshing problem \cite{faltinsen2000multidimensional,faltinsen2002asymptotic,faltinsen2010multimodal,lukovsky2012asymptotic,ansari2011two,gomez2013two}. Nevertheless, these relations are valid under reasonable assumptions for specific tank geometries, which poses a fundamental challenge in the multimodal method's application in arbitrary settings \cite{faltinsen2001adaptive}. It is, therefore, tempting to explore emerging data-driven tools to mitigate such problems. For example, as discussed in Section~\ref{sec:POD}, one could consider the POD procedure, which provides a systematic framework that yields a set of basis functions (accompanied by a sorting mechanism) from a set of snapshots.}

\textcolor{b}{
\subsection{Proper orthogonal decomposition: linear best-fit basis functions} \label{sec:POD}
In addition to boundary conditions, we might have archival data (i.e., snapshot fields) to help us construct the basis functions. PCA \cite{pearson1901principal} can be used to construct the basis functions that optimally represent the data. In 1933, a geometric representation of PCA has been proposed by Hotelling \cite{hotelling1933analysis}, and this concept has later become popular as empirical orthogonal functions (EOF) \cite{obukhov1947statistically} in environmental science, and POD \cite{aubry1991hidden} in the fluid dynamics community. The \emph{method of snapshots}\cite{sirovich1987turbulence} has been instrumental in the development of POD based approaches \cite{berkooz1993proper}. To compute the POD basis functions, let us assume that we have access to $m$ snapshots $\boldsymbol u(\boldsymbol x,t_{i})$ for $i=1,2, \dots, m$. A Reynolds decomposition-like expansion can be written as 
\begin{align}
\textcolor{rev}{\boldsymbol u(\boldsymbol x,t_{i}) = \bar{\boldsymbol u}(\boldsymbol x) + \boldsymbol \upsilon(\boldsymbol x,t_{i})}, 
\end{align}
where $\bar{\boldsymbol u}(\boldsymbol x)$ is a reference (e.g., ensemble mean) field, which can be obtained as
\begin{align}\label{mean}
\textcolor{rev}{\bar{\boldsymbol u}(\boldsymbol x) = \frac{1}{m}\sum_{i=1}^{m} \boldsymbol u(\boldsymbol x,t_{i})},
\end{align}
and the set of anomaly snapshots $\boldsymbol \upsilon(\boldsymbol x,t_{i})$ for $i=1,2, \dots, m$ can be defined as
\begin{align}\label{anomany}
\textcolor{rev}{\boldsymbol \upsilon(\boldsymbol x,t_{i})  = \boldsymbol u(\boldsymbol x,t_{i}) - \bar{\boldsymbol u}(\boldsymbol x)}. 
\end{align}
}

\textcolor{b}{
For clarity, we introduce the POD procedure for a scalar field (POD is normally applied to the velocity vector field). Let us denote $\upsilon(\boldsymbol x, t_j)$ a component of the anomaly velocity vector field (e.g., the $x$-component). A temporal correlation matrix $\textbf{A} = [\alpha_{ij}]$ can be constructed from these anomaly snapshots:
\begin{align}\label{pod}
  \alpha_{ij} = \int_{\Omega} \upsilon(\boldsymbol x, t_i) \upsilon(\boldsymbol x,t_j)d\boldsymbol x, \ \ \boldsymbol x \in \Omega,
\end{align}
where $\Omega$ is the spatial domain, and $i$ and $j$ refer to the snapshot indices. We define the $L^2$ inner product of two functions $f$ and $g$ as
\begin{align}
  \Big( f(\cdot),g(\cdot) \Big)= \int_{\Omega} f(\boldsymbol x)g(\boldsymbol x)d\boldsymbol x,
\end{align}
which yields $\alpha_{ij} = \Big( \upsilon(\boldsymbol x, t_i), \upsilon(\boldsymbol x, t_j)\Big)$ from Eq.~\ref{pod}. The data correlation matrix $\textbf{A} = [\alpha_{ij}]$ is a non-negative, symmetric $m \times m$ matrix, also known as the Gramian matrix of $\upsilon(\boldsymbol x, t_1)$, $\upsilon(\boldsymbol x, t_2)$, $\dots$, $\upsilon(\boldsymbol x, t_m)$. If we define the diagonal eigenvalue matrix $\boldsymbol\Lambda=\text{diag}[\lambda_1,....,\lambda_m]$ and a right eigenvector matrix
$\boldsymbol\Gamma = [\boldsymbol\gamma_1,....,\boldsymbol\gamma_m]$ whose columns are the corresponding eigenvectors of $\textbf{A}$, we can solve the following eigenvalue problem to obtain the optimal POD basis functions \cite{ravindran2000reduced}:
\begin{align}\label{eig}
  \textbf{A}\boldsymbol\Gamma  = \boldsymbol\Gamma \boldsymbol\Lambda.
\end{align}
In general, most of the subroutines for solving Eq.~\ref{eig} give $\boldsymbol \Gamma$ with all of the eigenvectors normalized to unity. The orthonormal POD basis functions for the anomaly field, $\upsilon$, can be thus calculated as follows:
\begin{align}
  \psi_i(\boldsymbol x) = \frac{1}{\sqrt{\lambda_i}}\sum_{k=1}^{m}\gamma_{i}^{k}\upsilon(\boldsymbol x,t_k),
\end{align}
where $\lambda_i$ is the $i^{th}$ eigenvalue, $\gamma_{i}^{k}$ is the $k^{th}$ component of the $i^{th}$ eigenvector, and $\psi_i(\boldsymbol x)$ is the $i^{th}$ POD mode. }

\textcolor{b}{
The eigenvalues are often stored in descending order for practical purposes, i.e., $\lambda_1 \geq \lambda_2 \geq ... \geq \lambda_m \geq 0$, and the eigenvectors are normalized in such a way that the basis functions satisfy the following orthonormality condition:
\begin{equation}
\Big( \psi_i,\psi_j\Big) = \left\{
\begin{aligned}
  1, \ \ i = j; \\
  0,  \ \ i \neq j.
  \end{aligned}
\right.
\end{equation}
Now, we can linearly represent the anomaly field variable $\upsilon(\boldsymbol x,t)$ using the POD modes as follows:
\begin{align}
  \upsilon(\boldsymbol x,t) = \sum_{i=1}^{r} a_i(t) \psi_i(\boldsymbol x),
\end{align}
where $a_i$ are the time-dependent (pseudo) modal coefficients, and $r$ is the total number of retained modes after the truncation, with $r\ll m$. These $r$ modes with the largest energy content correspond to the largest eigenvalues ($\lambda_1,\lambda_2,...,\lambda_r$). \textcolor{rev}{In general, adding more POD modes reduces the POD-ROM error.  We note, however, that this is not always true.  For example, adding POD modes that are polluted by numerical noise can actually decrease the POD-ROM error (see, e.g., the numerical investigation in Ref.\cite{giere2015supg}).} Often the value of $r$ is determined by using the relative information content (RIC) index \cite{gunzburger2002perspectives}, which is defined as
\begin{align}
  \mbox{RIC} = \frac{\sum_{i=1}^{r}\lambda_i}{\sum_{i=1}^{m}\lambda_i},
\end{align}
where $\mbox{RIC} = 1$ refers to a complete representation of the data snapshots. For example, if one records a set of snapshots from the field given by Eq.~\ref{eq:wave14} (i.e., the solution of the advection problem in Eq.~\ref{eq:wave}), let us say $m=100$ or more equally distributed snapshots between $t=0$ and $t=T$, the POD analysis could offer a perfect representation with $\mbox{RIC} = 1$ using only two retained modes ($r=2)$ since the underlying dynamics can be constructed by a linear superposition of two harmonics. Of course, that is not always the case, and $\mbox{RIC}$ becomes smaller than unity even if we retain a substantial number of modes, especially for turbulent flows.}

\textcolor{b}{This need for a large number of modes is one of the chief motivating factors for developing closure models to compensate the effects of the truncated modes in ROMs. 
However, it is believed that there is no separation of scales in turbulence, and therefore, most turbulent flow problems cannot be characterized by a high RIC index. Specifically, if there is no significant pattern in the evolution dynamics, there is a slow decay rate for the eigenvalues $\lambda_k$, and retaining only a few modes cannot capture the essential dynamics of turbulence. Thus, it is not surprising that many ROM practitioners have often demonstrated their proposed methodologies for problems that show somehow an underlying pattern (e.g., a shedding pattern in simulating the von Karman street). }

\textcolor{b}{This picture can be linked to the Kolmogorov barrier \cite{ahmed2020breaking}, where the linear reducibility (i.e., representing the underlying fluctuation field as a linear superposition/span of a finite/limited number of basis functions) is hindered. \emph{Modal expansions have an elliptic nature by construction, and using such tools for convection-dominated flows with higher degrees of hyperbolicity might often add another level of complexity when designing projection ROMs.} Then, a central question might arise about this data-driven procedure: \emph{Do we really get any benefit using the POD basis functions generated from prerecorded snapshots?} Although there might be a trade-off between storage, accuracy, and efficiency, the answer probably depends on the problem at hand. It might be a big \emph{yes} if there is an underlying pattern (e.g., limit cycles or quasi-periodic oscillations), and might be a \emph{no} if the flow is highly turbulent in a statistically non-equilibrium and chaotic state. In the latter case, one might consider a standard local discretization (e.g., finite difference/element/volume) method or a pseudo-spectral method (supported by the harmonics that satisfy the boundary conditions) without attempting to perform the POD procedure to compute a set of data-driven global basis functions.}

\textcolor{b}{For instance, the fast Fourier transform (FFT) provides an extremely efficient computational framework for models with such global basis functions without requiring any additional storage for precomputed or measured snapshots to generate a set of data-driven basis functions. The trade-off between accuracy and computational efficiency should always be considered carefully in generating data-driven models like Galerkin ROMs. The complexity of a typical right-hand side (RHS) computation of a pseudo-spectral solver becomes slightly bigger than $\mathcal{O}(n)$, where $n$ refers to the number of grid points. In contrast, the complexity of a typical Galerkin ROM is $\mathcal{O}(r^3)$ (there are also additional costs associated with, e.g., collecting and processing snapshots or solving an eigenvalue problem to generate a set of basis functions). Therefore, the Galerkin ROM becomes a computationally feasible approach \emph{if and only if} a few number of retained modes are utilized. As a rule of thumb, $r$ should be significantly less than the number of grid points in each direction for a canonical 3D problem (e.g., $r \ll 256$ for a $256^3$ problem). Otherwise, it would be hard to justify that the model is indeed \emph{reduced order}, since instead we could simply use the FFT algorithm to integrate the dynamical system equations in the harmonic space. Same arguments hold true for using a more flexible and convenient localized model, especially for problems with more complicated geometries (e.g., with the finite element, finite difference, or finite volume method, where the RHS can be obtained in $\mathcal{O}(n)$ computations).
}

\textcolor{b}{
\subsection{Leveraging ROR}
One of the major reasons for the inaccuracy of current ROMs in the numerical simulation of complex flows is the quality of ROR, which is the ability or inability of the ROM framework to represent the underlying complex dynamics. Specifically, in order to determine whether there is a valid ROR of the given system, we need to answer the following questions: (i) Is the ROM basis able to accurately approximate the dynamics? (ii) Is the Galerkin projection yielding an accurate ROM?}

\textcolor{b}{Furthermore, the issue of the selection of a convenient domain often comes into play. If the domain of influence is too small, there might be no good dynamical prediction. On the other hand, when it is too large, there could be too many uncorrelated events that have to be lumped in global modes. These uncorrelated events might work against the modeling accuracy since they often increase the deformation of modal expansion or degradation of the model representation. In other words, the domain should be as small as possible and as large as necessary. The correlation length might be a good initial reference scale to define the domain of interest. Dynamic mode adaptation, parameter-space-time domain partitioning, as well as smart clustering ideas have been explored, although we believe this topic is still in its infancy.}

\textcolor{b}{In practice, the construction of \emph{a good} low-order space is a cornerstone in projection-based ROM. That said, the representability of POD basis functions becomes questionable for non-stationary, strongly-evolving, and convection-dominated flows. Being a linear-based approach, POD might not be sufficient to describe nonlinear processes. More importantly, using a Galerkin projection based on elliptical ansatz for a hyperbolic problem could generate numerical oscillations. Moreover, the POD is optimal \emph{globally} in the sense that it minimizes the \emph{averaged} $L^2$ error across all the snapshot data. This raises the issue of modal deformation by the rapidly varying flow field state in such a way that the resulting modes are not representative of any of the system's states. Furthermore, since the POD modes are ranked based on their energy content, excursions in state spaces that contain a small amount of energy can be overlooked by POD even if these excursions might have significant impact on the dynamical evolution (see \citet{cazemier1998proper} for example). Similar scenarios arise for parameterized systems spanning a large parameter space when the system's behavior highly depends on the parameter value. Therefore, we devote the rest of this section to efforts aimed at enhancing the basis representability by either improving the offline construction stages or efficiently updating the ROM during online deployment.}

\textcolor{b}{One of the simplest approaches to improve the quality of the POD basis functions is to enrich the snapshot data matrix with \emph{extra} information. For example, in addition to the exact flow field data, the scaled difference between consecutive snapshots (i.e., the difference quotients) can be utilized such that the time derivative information is better represented in the resulting modes \cite{KV01,koc2021optimal}. Moreover, instead of collecting snapshot data at arbitrary time intervals and/or parameter values, more effective sampling techniques should be pursued. In Ref.~\cite{cui2015data}, a ROM is integrated into a Markov chain Monte Carlo (MCMC) framework, where the posterior distribution estimated by the MCMC algorithm is utilized to adaptively select the parameter values at which snapshots are evaluated.}

\textcolor{b}{In an effort towards the accurate identification of coherent structures from experimental or numerical data, a spectral POD approach has been developed \cite{sieber2016spectral}, and its relationship to DMD and resolvent analysis has been established \citep{towne2018spectral}. These studies use a technique that performs several PODs on individual frequencies obtained from Fourier-transformed windows of snapshot data. Thus, the modes we get for each frequency correspond to a coherent frequency domain structure. If an analysis of the eigenvalue spectrum for each frequency reveals coherent structures, it can indicate that there is a physical process which is occurring at that frequency. Hence, this becomes a useful data analysis tool on top of providing orthogonal modes for ROMs. In addition, transported snapshots POD approaches \citep{reiss2018shifted,mendible2020dimensionality} have been introduced for convection-dominated transport systems. In particular, these studies use a shifting operator on the snapshots (requiring interpolation on unstructured grids and some knowledge of the transport speed) to allow POD or DMD to (more) efficiently approximate advective systems. In their recent works, \citet{mendible2021data} employed an unsupervised traveling wave identification with shifting and truncation (UnTWIST) algorithm \cite{mendible2020dimensionality} to discover moving coordinate frames into which the data are shifted, thus overcoming limitations imposed by the underlying translational invariance and allowing for the application of traditional dimensionality reduction techniques. \citet{etter2020online} proposed a novel online adaptive basis refinement mechanism for efficiently enriching the trial basis in a manner that ensures convergence of the ROM to the FOM.}

\textcolor{b}{Localization methods have been successfully pursued to mitigate the modal deformation of the POD basis by partitioning the state space \cite{amsallem2012nonlinear,washabaugh2012nonlinear,peherstorfer2014localized,taddei2015reduced}, time domain \cite{ijzerman2000signal,dihlmann2011model,borggaard2007interval,san2015principal,Babaee2016ptrs,chaturantabut2017temporal,ahmed2018stabilized,ahmed2019memory,ahmed2020breaking}, physical domain \cite{maday2002reduced,lovgren2006reduced,iapichino2012reduced,eftang2013port}, or parameter space \cite{eftang2012parameter,moosavi2015efficient,hess2019localized} using multiple local, piecewise affine subspace approximations instead of a single global approximation. These partitioning or time varying approaches work by parsing the available snapshot data into a few overlapping or non-overlapping groups (e.g., based on solution value, time, parameter, geometry, or component) and applying standard modal decomposition techniques (e.g., POD) for each region separately. This eventually yields a library of compact ROMs, each suitable for a specific region and/or dynamics, and interpolation methods can be utilized when the region of interest does not exist in the available library.}

\textcolor{b}{In this context, clustering techniques can be also utilized to effectively perform such partitioning. Indeed, cluster-based reduced order models (CROMs) have been proposed to tackle some of the potential pitfalls of classical GROMs (e.g., the mismatch between the modal expansion approach and the underlying dynamics, see \citet{noack2016snapshots}). CROMs start by sorting the snapshot data into a small number of clusters (e.g., using k-means approach) with centroids being the representative states in each cluster. Conceptually, this is similar to coarse-graining the state-space (or generally the feature-space) into centroidal Voronoi tessellation (CVT) generators \cite{burkardt2006centroidal}. The transition dynamics between these centroids can be modeled as a probabilistic Markov model \cite{kaiser2014cluster,kaiser2017cluster,li2020cluster} or a deterministic–stochastic network model \cite{fernex2019cluster,li2021cluster,fernex2021cluster}.} 

\begin{figure}[ht]
\centering
\includegraphics[width=0.95\linewidth]{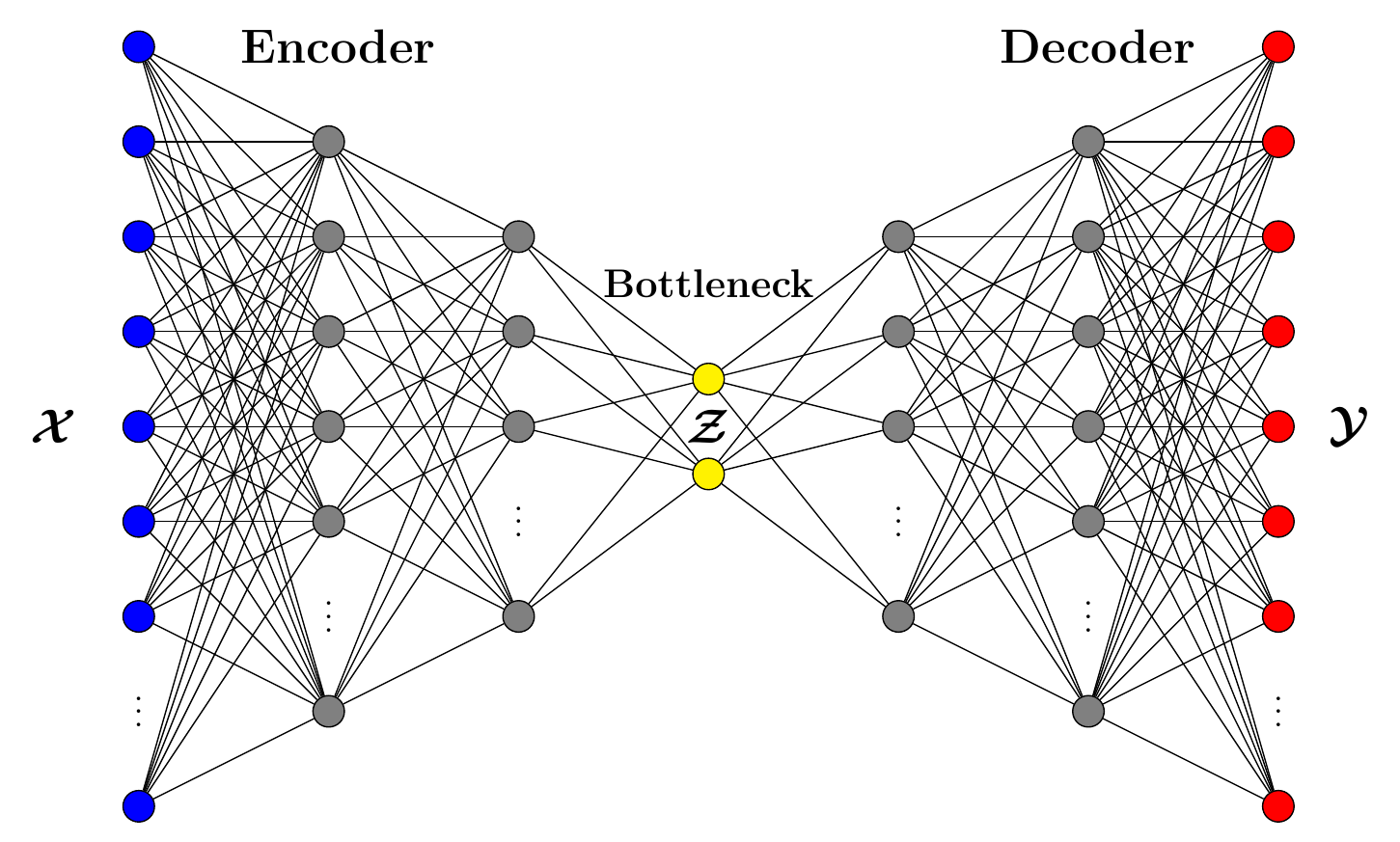}
\caption{A schematic diagram for an autoencoder (AE) for latent space construction, where the input $\mathcal{X}$ is the full field, the output $\mathcal{Y}$ designates its reconstruction, and the bottleneck $\mathcal{Z}$ represents the latent space (compressed) variables. }
\label{fig:AE}
\end{figure}

\textcolor{b}{As highlighted in Section~\ref{sec:POD}, POD provides an efficient way to compress data and explain the variance of the data better than any other linear combination \cite{jolliffe2002principal}. Indeed, from a linear algebra perspective, it can often be formulated as a singular value decomposition, providing an optimal low-rank matrix approximation. This can leverage highly performant and scalable algorithms to handle extremely large datasets, benefiting from the rich legacy of linear algebra investigations. From a statistical point of view, this orthogonal projection provides \emph{linearly uncorrelated} features. However, it cannot reveal nonlinear correlations in the data.
In contrast, \emph{manifold learning} (or representation learning) techniques aim at accounting for such nonlinear correlations to further reduce the dimensionality of the problem. The generalizations of PCA to nonlinear settings often define a curve in the latent space which minimizes the mean squared error of all variables. Yet, the smoothness of the curve can be varied by the method. For example, an autoassociative or autoencoding neural network model \cite{kramer1991nonlinear,hsieh2001nonlinear,hsieh2009machine} and a kernel PCA \cite{scholkopf1998nonlinear} are two successful approaches of such a nonlinear PCA (NLPCA) framework. We refer the reader to recent works \cite{otto2019linearly,fukami2020convolutional,agostini2020exploration} for excellent discussions on autoencoder technology in fluid dynamics (see Figure~\ref{fig:AE}). We also note that other nonlinear dimensionality reduction techniques, such as principal curves \cite{hastie1989principal}, locally linear embedding \cite{roweis2000nonlinear}, isomap \cite{tenenbaum2000global} and self-organized map \cite{kohonen1982self} approaches, can also be regarded as a discrete version of NLPCA.}

\textcolor{rev}{It is worth noting that during the modal truncation step (reducing the dimensionality of the system), dependencies among retained and discarded modes generally yield inaccurate results if the closure problem is not addressed. In this regard, persistent homology (PH) \cite{edelsbrunner2000topological,edelsbrunner2008persistent,edelsbrunner2014persistent,zomorodian2008localized,zomorodian2005computing} provides a delicate balance between data simplification and intrinsic structure extraction. PH is a tool in topological data analysis that aims at studying and extracting the features that persist across multiple scales by casting the multi-scale organization into a mathematical formalism. In particular, PH measures the lifetime of intrinsic topological features using a filtration process to distinguish between the long-lived features and the short-lived ones (which are considered topological noise) \cite{otter2017roadmap}.}

\textcolor{rev}{However, the application of PH has been largely dedicated to qualitative data classification and analysis, and its utility for quantitative modeling and prediction (including ROM) is scarce \cite{wang2016object}. PH has been utilized to characterize the time series from dynamical systems based on topological features that appear in the solution manifold or attractor \cite{pereira2015persistent,garland2016exploring,maletic2016persistent}. \citet{rieck2015persistent} used PH as an evaluation tool to compare the performance of different dimensionality reduction algorithms (e.g., PCA, and isomap mentioned earlier). One of the major challenges of employing PH is its prohibitive computational cost (for the worst-case scenario). To increase the PH efficiency, \citet{moitra2018cluster} utilized a clustering technique to represent similar groups of data points with their cluster centroid and applied PH onto these clusters. }

\section{First Principle Galerkin Method} \label{sec:Galerkin}

Finite dimensional low-order models routinely arise when we apply Galerkin type projection techniques to infinite dimensional PDE models \cite{lakshmivarahan2008relation,lakshmivarahan2008structure,wang2009relation}. We formalize the model reduction problem for fluid flow systems, considering a generic prognostic equation as follows:
\begin{equation}
    \dfrac{\partial \mathbf{u}}{\partial t} = \mathcal{\boldsymbol F}(\mathbf{u};\mathbf{x},t),
\end{equation}
where $\mathbf{u}$ denotes the discrete approximation of a three-dimensional (3D) dependent variable (e.g., density, velocity, temperature, moisture); $\mathbf{x}$ denotes the independent spatial variables (e.g., latitude, longitude, and height); and $\mathcal{\boldsymbol F}$ defines the model's dynamical core (e.g., semi-discretized PDEs representing mass, momentum, and energy conservation), all written in vector-form. More specifically, we explore the autonomous dynamics case for a specific quantity of interest $\mathbf{u}$, with $\mathcal{\boldsymbol F}$ being decomposed into linear $\mathcal{\boldsymbol L}$ and nonlinear $\mathcal{\boldsymbol N}$ operators as
\begin{equation}
    \dfrac{\mathrm{d} \mathbf{u}}{\mathrm{d}  t} = \mathcal{\boldsymbol L}\mathbf{u} +  \mathcal{\boldsymbol N}(\mathbf{u}), \label{eq:uode}
\end{equation}
where $\mathbf{u} \in \mathbb{R}^n$, where $n$ is the number of degrees of freedom in the spatial discretization. Considering the Navier-Stokes equations as a typical mathematical framework for fluid flow modeling, we highlight that the linear and nonlinear operators often represent the diffusive and convective effects, respectively. 



In order to build the projection-based ROM, the solution $\mathbf{u}$ is approximated in a low-dimensional affine subspace of dimension $r$ via the Galerkin ansatz as follows:
\begin{equation}
\mathbf{u}(t) \approx \bar{\mathbf{u}} + \boldsymbol{\Psi} \ba(t), \label{eq:urom}
\end{equation}
where $\bar{\mathbf{u}} \in  \mathbb{R}^n$ is a reference solution representing the affine offset, $\boldsymbol{\Psi} \in \mathbb{R}^{n \times r}$ denotes the trial basis, and $\ba(t) \in \mathbb{R}^r$ is the vector of reduced (generalized) coordinates, also called modal weights or coefficients. The reference solution $\bar{\mathbf{u}}$ as well as the basis $\boldsymbol{\Psi}$ are constructed during an offline stage from a collection of FOM evaluations (called snapshots). Without loss of generality, we suppose that the time-averaged field defines the reference solution, $\bar{\mathbf{u}}$,  and the basis $\boldsymbol{\Psi}$ is constructed using the POD technique. Then, the low-rank approximation given by Eq.~\ref{eq:urom} is substituted into Eq.~\ref{eq:uode} and an inner product with a test basis is performed to yield a system of ODEs for the unknown modal coefficients, $\ba(t)$. In Galerkin projection-based ROM (GROM), the test basis is chosen to be the same as the trial basis. We make use of the orthonormality property (i.e., $\boldsymbol{\Psi}^{\top} \boldsymbol{\Psi} = \boldsymbol{I}_r$, where $\boldsymbol{I}_r$ is the $r\times r$ identity matrix and the superscript $\top$ denotes the matrix transpose, assuming a Euclidean state space) as follows:
\begin{equation}
    \boldsymbol{\Psi}^{\top}\dfrac{\mathrm{d} }{\mathrm{d}  t} \left(\bar{\mathbf{u}} + \boldsymbol{\Psi} \ba\right)= \boldsymbol{\Psi}^{\top}\mathcal{L} \left(\bar{\mathbf{u}} + \boldsymbol{\Psi} \ba\right) +  \boldsymbol{\Psi}^{\top}\mathcal{N}\left(\bar{\mathbf{u}} + \boldsymbol{\Psi} \ba\right). \label{eq:grom0}
\end{equation}
Since both $\bar{\mathbf{u}}$ and $\boldsymbol{\Psi}$ are considered time-independent, Eq.~\ref{eq:grom0} reduces to
\begin{equation}
    \boldsymbol{\Psi}^{\top}\boldsymbol{\Psi} \dfrac{\mathrm{d} \ba }{\mathrm{d}  t} = \boldsymbol{\Psi}^{\top}\mathcal{L} \bar{\mathbf{u}} + \boldsymbol{\Psi}^{\top}\mathcal{L} \boldsymbol{\Psi} \ba +  \boldsymbol{\Psi}^{\top}\mathcal{N}\left(\bar{\mathbf{u}} + \boldsymbol{\Psi} \ba\right). \label{eq:grom1}
\end{equation}
Note that $\boldsymbol{\Psi}^{\top}\mathcal{L} \bar{\mathbf{u}}$ and $\boldsymbol{\Psi}^{\top}\mathcal{L} \boldsymbol{\Psi}$ can be precomputed during the offline construction stage, reducing the online computational cost of evaluating the first two terms on the right-hand side to $O(r)$ independent of the FOM dimension, $n$. However, generally speaking, computing the third term representing the system's nonlinearity depends on $n$, limiting the computational benefit of ROM. In order to mitigate this limitation, hyperreduction approaches have been developed to relieve this dependency on $n$, by \emph{approximating}, rather than \emph{evaluating}, the nonlinear term in a reduced order subspace \cite{cstefuanescu2014comparison,dimitriu2017comparative,yano2019discontinuous}. Examples of hyperreduction include the empirical interpolation method (EIM) \cite{barrault2004empirical}, its discrete version (DEIM) \cite{chaturantabut2009discrete,chaturantabut2010nonlinear}, the gappy POD \cite{everson1995karhunen,bui2004aerodynamic,carlberg2013gnat}, and the missing point estimation (MPE) \cite{astrid2008missing,zimmermann2016accelerated}, where the approximation is performed using sampling techniques. On the other hand, tensorial ROM can benefit from the quadratic (or generally polynomial) nonlinearity, which is ubiquitous in fluid flow systems, to rewrite Eq.~\ref{eq:grom1} as follows:
\begin{equation}
    \dfrac{\mathrm{d} \ba }{\mathrm{d}  t} = \mathbf{\mathfrak{B}} +  \mathbf{\mathfrak{L}} \ba +  \ba^\top \mathbf{\mathfrak{N}} \ba, \label{eq:grom2}
\end{equation}
where the vector $\mathbf{\mathfrak{B}}$, the matrix $\mathbf{\mathfrak{L}}$, and the tensor $\mathbf{\mathfrak{N}}$ are precomputed during the offline stage, reducing the computational cost of solving the GROM defined in Eq.~\ref{eq:grom2} to $O(r^3)$ in case of quadratic nonlinearity (which is the case for the Navier-Stokes equations). When the true underlying dynamics of the system are non-polynomial, \emph{lifting} transformations can be exploited to yield a finite-dimensional coordinate representation in which the system dynamics have quadratic structure \cite{gu2011qlmor,benner2015two,benner2018mathcalh_2,kramer2019nonlinear}. Although such transformation is not universally guaranteed, a large class of smooth nonlinear systems that appear in engineering applications (e.g., elementary functions like exponential and trigonometric functions or polynomials) can be equivalently lifted to quadratic form.

\textcolor{b}{
\subsection{POD Galerkin projection: Burgers equation} \label{sec:burg}
To illustrate the POD Galerkin approach for flow systems with quadratic nonlinearity, let us consider the Burgers equation
\begin{align}\label{eq:burgers}
  \frac{\partial u}{\partial t} + u\frac{\partial u}{\partial x} = \nu \frac{\partial^2 u}{\partial x^2},
\end{align}
which is often used as a simplified prototype by fluid dynamicists. Using the POD procedure outlined in Section~\ref{sec:POD}, we can define the $u(x,t)$ field as a linear superposition of the mean field and the POD basis functions,
\begin{align}
 u(x,t) = \bar{u}(x) + \sum_{i=1}^{r} a_i(t) \psi_i(x),
\end{align}
and substitute this approximation of our field variable into Eq.~\ref{eq:burgers}. Once we perform an orthonormal Galerkin projection, the resulting dynamical system for $a_k(t)$ can be written as
\begin{equation} \label{eq:rom1}
  \frac{d a_k}{d t} = \mathfrak{B}_{k} + \sum_{i=1}^{r} \mathfrak{L}^{i}_{k}a_{i} + \sum_{i=1}^{r}\sum_{j=1}^{r} \mathfrak{N}^{ij}_{k}a_{i}a_{j},
\end{equation}
where
\begin{eqnarray}
  & & \mathfrak{B}_{k} = \big( \nu \frac{\partial^2 \bar{u} }{\partial x^2} - \bar{u}\frac{\partial \bar{u}}{\partial x}, \psi_{k} \big) , \nonumber \\
  & & \mathfrak{L}^{i}_{k} = \big( \nu \frac{\partial^2 \psi_i }{\partial x^2} - \bar{u}\frac{\partial \psi_i}{\partial x} - \psi_i \frac{\partial \bar{u}}{\partial x}, \psi_{k} \big)  , \nonumber\\
  & &  \mathfrak{N}^{ij}_{k} = \big( - \psi_i\frac{\partial \psi_j}{\partial x}, \psi_{k} \big). \label{eq:roma7}
\end{eqnarray}
This tensorial system consists of $r$ coupled ODEs and it is often written as Eq.~\ref{eq:grom2}, where $\boldsymbol a$ is the vector of unknown coefficients $a_k(t)$, $k=1,2, \dots, r$, $\mathbf{\mathfrak{B}}$ is a scaling vector coming from the reference mean field with entries $\mathfrak{B}_{k}$, $\mathbf{\mathfrak{L}}$ is an $r \times r$ matrix with entries $\mathfrak{L}^{i}_{k}$ for the contribution stemming from the linear viscous term, and $\mathbf{\mathfrak{N}}$ is an $r \times r \times r$ tensor with entries $\mathfrak{N}^{ij}_{k}$ arising from the nonlinear advection term, $1 \leq i, j, k \leq r$. In this tensorial form, the corresponding model coefficients $\mathfrak{B}_{k}$, $\mathbf{\mathfrak{L}}$, and  $\mathbf{\mathfrak{N}}$ are \emph{precomputed} from the available snapshots. Alternatively, there are a number of \emph{online} approaches where we can compute the nonlinear part using hyperreduction or principled sampling strategies to approximate the full nonlinear state from a small number of measurement or collocation points \cite{everson1995karhunen,barrault2004empirical,willcox2006unsteady,yildirim2009efficient,chaturantabut2010nonlinear,drmac2016new}. \citet{cstefuanescu2014comparison} performed a comparative study between the direct (online) and tensorial (precomputed) methods. Moreover, \citet{karasozen2021structure} recently discussed structure preserving ROMs and compared the direct and tensorial POD approaches.}


\textcolor{b}{
\subsection{Projection based ROMs} \label{sec:gap}
Computational models for the Navier-Stokes equations could make a tremendous impact in critical applications, such as the biomedical and engineering applications that we describe next. Despite their enormous potential, FOMs have not fully transitioned to engineering practice. The main roadblock is the extraordinary computational cost incurred by computational models in many applications. For example, although preliminary studies of aortic dissections showed that uncertainties in the geometry and inflow conditions have a fundamental role, performing an {\it uncertainty quantification} study requires a huge number of computational model runs. Similarly, performing a {\it shape optimization} study to determine the optimal vascular configuration for the total cavopulmonary connection surgery requires again many computational model runs. Also, in renewable energy applications, performing {\it data assimilation} to incorporate the available observations in the control of wind-power production requires numerous model runs.}

\textcolor{b}{Since running current computational models hundreds and thousands of times can take days and weeks on high performance computing (HPC) platforms, a brute force computational approach for these biomedical and engineering applications is simply not possible. Therefore, what is needed is a modeling strategy that allows model runs that take minutes to hours on a laptop.}

\textcolor{b}{For structure-dominated systems, ROMs can decrease the FOM computational cost by {\it orders of magnitude}. ROMs are (extremely) low-dimensional models that are trained (constructed) from available data. As explained in Section~\ref{sec:Galerkin}, in an offline phase, the FOM is run for a few parameters values to construct a low-dimensional (e.g., $10$-dimensional) ROM basis $\{ \psi_{1}, \ldots, \psi_{10}\}$, which is used to build the ROM:
\begin{eqnarray}
	\frac{d\ba}{dt} =  \bff(\ba), 	\label{eqn:eqn-rom}
\end{eqnarray}
where $\ba$ is the vector of coefficients in the ROM approximation $\sum_{i=1}^{10} a_{i}(t) \psi_{i}(\bx)$ of the variable of interest and $\bff$ comprises the ROM operators (e.g., vectors, matrices, and tensors) that can be preassembled from the ROM basis in the offline phase. In the online phase, the low-dimensional ROM given by Eq.~\ref{eqn:eqn-rom} is then used for parameters values that are {\it different} from those used in the training stage. Since ROM is low-dimensional (10-dimensional), its computational cost is orders of magnitude lower than the FOM cost.
Thus, for the biomedical and engineering applications described above, ROMs appear as a natural alternative to the prohibitively expensive FOMs.}


\textcolor{b}{Unfortunately, current ROMs cannot be used in clinical and engineering practice, since they would require too many modes (degrees of freedom). For example, to capture all the geometric scales in aortic dissection, one might need hundreds or even thousands of ROM modes (e.g., see Table~\ref{table:modes}). Similarly, to cope with the high Reynolds number in the wind farm optimization, a large number of ROM modes are necessary. Thus, although ROMs decrease the FOM computational cost by orders of magnitude, they are still too expensive: Current ROMs cannot be run in minutes or hours on a laptop and thus cannot be used easily in clinical and engineering practice.}

\begin{table*}[htbp]
    \centering
    \caption{A non-exhaustive list illustrating energy characteristics and range of the number of retained modes $r$ for $m$ given snapshots.}
    \begin{tabular}{p{0.25\linewidth}p{0.23\linewidth}p{0.1\linewidth}p{0.07\linewidth}p{0.3\linewidth}}
    \hline\noalign{\smallskip}
    Study & Problem & $r$ & $m$ & Comment\\
    \noalign{\smallskip}\hline\noalign{\smallskip}
    \citet{osth2014need}  & Ahmed body & 10 -- 100 & 2000 & The first 500 modes resolve 60\% of the kinetic energy.\\
    \citet{san2015principal}  & Marsigli flows & 6 -- 30 & 400 & The first 30 modes resolve 90\% of the kinetic energy.\\
    \citet{rahman2019dynamic}  & Quasigeostrophic flows & 10 -- 80 & 400 & The first 50 modes resolve 80\% of the kinetic energy.\\ 
    \citet{ballarin2016fast}  & Hemodynamics & 50 & 400 & O(10--100) modes are required to obtain a reliable approximation.\\ 
    \citet{verhulst2014large}  & Wind farm & N/A & 7200 & 430 POD modes are required to capture 80\% of
    the total energy.\\
    \citet{shah2014very}  & Atmospheric boundary layer & N/A & 2500 & 500 POD modes are required to capture 80\% of
    the total energy. \\
    \citet{zhang2020characterizing}  & Atmospheric boundary layer & N/A & 5000 & 2000 POD modes are required to capture 80\% of
    the total energy.\\
    \noalign{\smallskip}\hline
    \end{tabular}
    \label{table:modes}
\end{table*}

\begin{figure}[ht]
\centering
\includegraphics[width=0.80\linewidth]{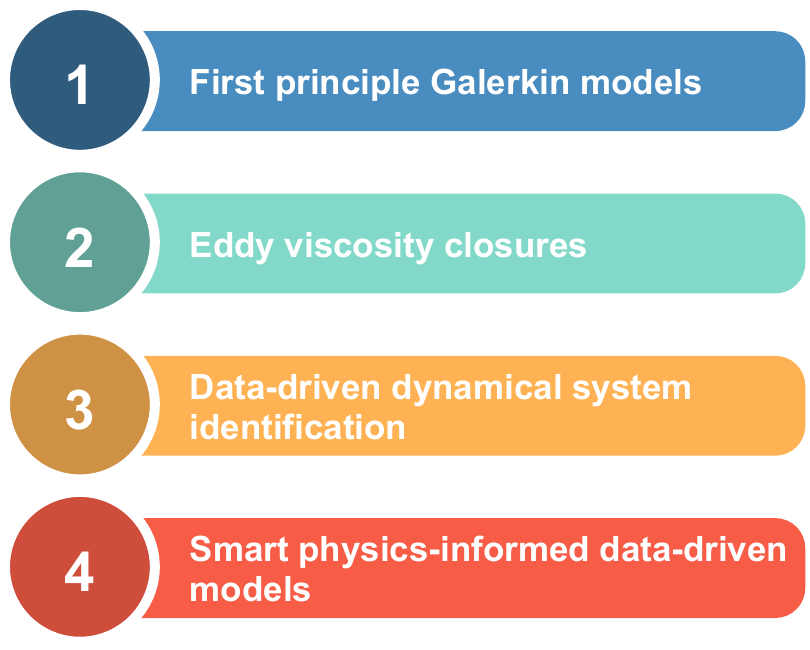}
\caption{Evolution of the ROM approaches.}
\label{fig:steps}
\end{figure}

\textcolor{b}{With the evolution of ROM approaches outlined in Figure~\ref{fig:steps} in mind, the ROM community is at a \emph{crossroads}. On the one hand, current ROMs can be used for academic test problems for which a handful of ROM modes can model simple dynamics with substantial success. On the other hand, realistic, complex flows require high-dimensional ROMs that cannot be used in clinical and engineering practice. What is needed is low-dimensional, \emph{efficient} ROMs that are \emph{accurate} so that they can be utilized in such vital applications.}

\textcolor{b}{One of the main reasons for the notorious inaccuracy of current ROMs in complex clinical and engineering settings is the {\it drastic ROM truncation}:
Instead of using many (e.g., 100) ROM modes $\{ \psi_{1}, \ldots, \psi_{100}\}$, current ROMs use only a handful of ROM modes $\{ \psi_{1}, \ldots, \psi_{10}\}$ to ensure a low computational cost. This drastic truncation yields acceptable results in simple, academic test problems, but produces inaccurate results in practical clinical and engineering settings~\cite{osth2014need}.
Thus, for accurate results, the {\it ROM closure problem} needs to be solved:
One needs to model the effect of the discarded ROM modes $\{ \psi_{11}, \ldots, \psi_{100}\}$ on the ROM dynamics, i.e., on the time evolution of resolved ROM modes $\{ \psi_{1}, \ldots, \psi_{10}\}$:
\begin{eqnarray}
	\frac{d\ba}{dt}
	=
	\bff(\ba)
	+
	\text{Closure}(\ba),
	\label{eqn:eqn-rom-closure}
\end{eqnarray}
where $\text{Closure}(\ba)$ is a low-dimensional term that models the effect of the discarded ROM modes $\{ \psi_{11}, \ldots, \psi_{100}\}$ on $\{ \psi_{1}, \ldots, \psi_{10}\}$. The closure term is also known as \emph{unresolved tendency}, or \emph{model error} in different disciplines.}

\textcolor{b}{The closure problem is prevalent in numerical simulation of complex systems. For example, classical numerical discretization of turbulent flows (e.g., finite element or finite volume  methods), inevitably takes place in the {\it under-resolved regime} (e.g., on coarse meshes) and requires closure modeling (i.e., modeling the sub-grid scale effects). In computational fluid dynamics (CFD), e.g., large eddy simulation (LES), there are hundreds (if not thousands) of closure models~\cite{sagaut2006large}. This is in stark contrast with reduced order modeling, where only relatively few ROM closure models have been investigated. The reason for the discrepancy between ROM closure and LES closure is that the latter has been mostly built around physical insight stemming from Kolmogorov's statistical theory of turbulence (e.g., the concept of eddy viscosity), which is generally posed in the Fourier setting~\cite{Pop00,sagaut2006large}. Much of this physical insight is not generally available in a ROM setting. Thus, current ROM closure models have been deprived of this powerful methodology that represents the core of most LES closure models. To  construct low-dimensional and efficient ROMs that are \emph{accurate}, a set of principled, mathematical and/or data-driven ROM closure modeling strategies need to be utilized. In Section~\ref{sec:closure}, we survey the main types of closure models developed in the reduced order modeling community.}

\section{Closure Modeling} \label{sec:closure}

Although the solution of Eq.~\ref{eq:grom2} becomes independent of the FOM dimension $n$, the cubic scaling with respect to $r$ hurts the turnaround of such ROMs. This is especially true for fluid flows of practical interest (e.g., turbulent and convection-dominated flows), where the FOM solution manifold is characterized by a large and slowly decaying Kolmogorov $n$-width \cite{kolmogoroff1936uber,pinkus2012n}. Thus, a large number of modes are required to maintain the solution accuracy, resulting in excessive computational overhead, which may even exceed the FOM computational cost. Therefore, in these complex settings, ROM will always incur a degree of under-resolution by sacrificing some degree of accuracy for the sake of computational efficiency. This under-resolution has direct and indirect consequences. The direct outcome is the projection error affecting the Galerkin ansatz (Eq.~\ref{eq:urom}), where some of the underlying flow features are lost. The indirect ramifications are related to the nonlinearity of the system, implying that the discarded modes indeed interact with the retained ones. By performing severe modal truncation (remember, computational efficiency is a priority!), we suppress these interactions and Eq.~\ref{eq:grom2} no longer captures the projected trajectory, decreasing the solution accuracy \cite{ahmed2020long}.

\textcolor{rev}{To illustrate the above discussion, consider a state variable $\mathbf{u}(t) \in \mathbb{R}^n$, which can be \emph{exactly} written as a superposition of $n$ basis functions as $\mathbf{u}(t) = \boldsymbol{\Psi} \ba(t) + \boldsymbol{\Phi} \boldsymbol{b}(t)$, where $\boldsymbol{\Psi} \in \mathbb{R}^{n\times r}$ and $\boldsymbol{\Phi} \in \mathbb{R}^{n\times (n-r)}$ represent the modes to be retained and truncated, respectively, and $\ba(t) \in \mathbb{R}^r$ and $\boldsymbol{b}(t)\in \mathbb{R}^{(n-r)}$ are the corresponding time-dependent coefficients. A Galerkin projection of the governing equations onto $\boldsymbol{\Psi}$ and  $\boldsymbol{\Phi}$ yields the following:
\begin{equation}
    \frac{d}{dt} \begin{bmatrix} \ba \\ \boldsymbol{b} \end{bmatrix} =  \begin{bmatrix} \bff_{\ba}(\ba,\boldsymbol{b}) \\ \bff_{\boldsymbol{b}}(\ba,\boldsymbol{b}) \end{bmatrix}. \label{eq:mz1}
\end{equation}
}

\textcolor{rev}{We note that Eq.~\ref{eq:mz1} is an exact representation of the system's dynamics. In reduced order modeling, we are only interested in the resolved part of the dynamics, which can be written as
\begin{equation}
    \frac{d\ba}{dt} = \bff_{\ba}(\ba,\boldsymbol{b}). \label{eq:mz2}
\end{equation}
Nevertheless, Eq.~\ref{eq:mz2} is not practical because its solution requires the knowledge of the unresolved variable, $\boldsymbol{b}$. In a classic truncated ROM, it is often assumed that $\bff_{\ba}(\ba,\boldsymbol{b}) = \bff_{\ba}(\ba,\boldsymbol{0}) = \bff(\ba)$. However, for nonlinear cases, this relation does not hold (i.e., $\bff_{\ba}(\ba,\boldsymbol{b}) \neq \bff_{\ba}(\ba,\boldsymbol{0})$).}

\begin{figure*}[ht]
\centering
\includegraphics[width=0.95\linewidth]{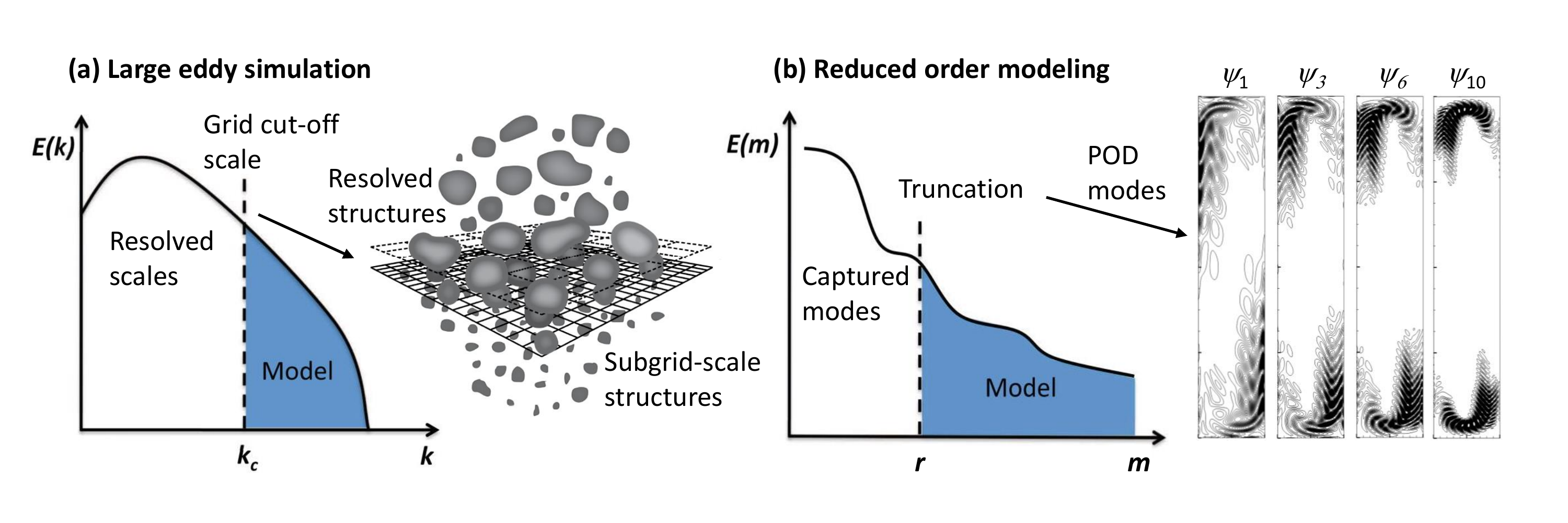}
\caption{Closure modeling analogy between LES and ROM, where higher values of $k$ and $m$ refer to smaller scales (adapted from \cite{akhtar2012new,san2014basis,azadani2015large}). Shaded areas in energy spectra represent the discarded scales that must be modeled.}
\label{fig:les}
\end{figure*}

Following the Kolmogorov hypotheses \cite{kolmogorov1941local,kolmogorov1991local} from turbulence modeling and assuming an analogy between POD and Fourier modes (see Figure~\ref{fig:les}), it is commonly agreed in the ROM community that the first POD modes resolve the large energy-containing flow scales, while the last modes correspond to the low-energy dissipative scales. Indeed, this analogy has been demonstrated theoretically and numerically for different flow scenarios (e.g., flow over a cylinder \cite{noack2003hierarchy} and a turbulent flow past a backward-facing step \cite{couplet2003intermodal}). Thus, truncating the low-energy scales is believed to result in a pile-up of energy levels, leading to solution instability. We also highlight that this argument has been recently the focus of scientific revisits. For instance, \citet{grimberg2020stability} state, using mathematical arguments and analogies from finite element analysis, that the solution instabilitiy observed in most studies dealing with GROM is a byproduct of the Galerkin projection step. Moreover, they show that a ROM based on Petrov-Galerkin projection, where the test basis differs from the trial basis, yields more accurate and stable solution than standard Galerkin projection. However, the test basis needs to be updated at each iteration and time step, increasing the computational complexity of the resulting ROM. As a highly promising approach, an adjoint Petrov-Galerkin method for nonlinear model reduction has been recently put forth by \citet{parish2020adjoint}. Rather than constructing a low-dimensional subspace for the entire state space in a monolithic fashion, \citet{hoang2021domain} recently proposed a dynamic methodology to construct separate subspaces for the different subdomains. Although the Petrov-Galerkin projection \cite{xiao2013non} could mitigate some of the challenges the Galerkin ROMs have to face in the under-resolved simulation of turbulent flows, we limit ourselves to Galerkin projection-based ROMs in the current review, where closure models have been mainly developed to improve the solution accuracy and stability properties. The major aim of closure models is to make up for the effects of discarded modes onto the dynamics of resolved modes. Specifically, the objective is to modify Eq.~\ref{eq:grom2} to correctly resolve the time dynamics of $\boldsymbol{\Psi}$: 
\begin{equation}
    \dfrac{\mathrm{d} \ba }{\mathrm{d}  t} = \mathbf{\mathfrak{B}} +  \mathbf{\mathfrak{L}} \ba +  \ba^\top \mathbf{\mathfrak{N}} \ba + \mathbf{\mathfrak{C}}, \label{eq:gromC}
\end{equation}
where $\mathbf{\mathfrak{C}}$ represents the closure model that needs to be determined.

The closure problem has historical roots in CFD, in particular in turbulence modeling, including the Reynolds averaged Navier-Stokes (RANS) and LES. In contrast, there are relatively fewer closure models that have been investigated in a ROM context. In general, the ROM closure modeling approaches can be classified into (1) functional (phenomenological), which use physical insights to postulate a model form for the closure term (e.g., a dissipative term) and (2) structural (mathematical), which often rely on filtering techniques to reveal the closure term without using any physical assumptions or additional phenomenological arguments.
We emphasize that placing a given ROM closure model in one of these categories is not always straightforward, as these categories sometimes overlap.
We also refer readers to the unified exposition of several mean field modeling ideas \cite{johnston2012mean} as well as other closure techniques for probability density function (PDF) \cite{brennan2018data} or moment closures for kinetic theories \cite{levermore1996moment}.

\begin{table*}[htbp]
    \centering
    \caption{A chronological list of key contributions to ROM closure modeling.}
    \begin{tabular}{p{0.05\linewidth}p{0.32\linewidth}p{0.60\linewidth}}
    \hline\noalign{\smallskip}
    Year & Study & Key Contribution \\
    \noalign{\smallskip}\hline\noalign{\smallskip}
    1988 & \citet{aubry1988dynamics} & First closure model: global eddy viscosity modeling\\
    1994 & \citet{rempfer1994dynamics} & Linear modal eddy viscosity closure\\
    1995 & \citet{selten1995efficient} & Time averaging closure modeling\\
    1997 & \citet{selten1997statistical} & A statistical closure of a barotropic model\\
    1998 & \citet{cazemier1998proper} & Penalty term closure model based on energy conservation principles \\
    2003 & \citet{couplet2003intermodal}  & Guidelines for modeling unresolved modes in POD–Galerkin models \\
    2004 & \citet{sirisup2004spectral} & Spectral viscosity closure for POD models \\
    2008 & \citet{noack2008finite} & Finite time thermodynamics and ensemble averaging closure models \\
    2009 & \citet{bergmann2009improvement,bergmann2009enablers} & \textcolor{rev}{Residual-based variational multiscale POD} \\
    2011 & \citet{borggaard2011artificial} & First numerical analysis of closure models: artificial viscosity model \\
    2011 & \citet{akhtar2012new} & Nonlinear eddy viscosity model based on the Frobenius norm of the Jacobian \\
    2011 & \citet{wang2011two} & Two-level discretization model \\
    2012 & \citet{wang2012proper} & Eddy viscosity variational multiscale and dynamic Smagorinsky closures \\
    2013 & \citet{balajewicz2013low} & Subspace calibration using the Navier-Stokes equations \\
    2013 & \citet{cordier2013identification} &  Proof of global boundedness of nonlinear eddy viscosity closures\\
    2014 & \citet{osth2014need} & $\sqrt{K}$-scaled eddy viscosity concept\\
    2014 & \citet{iliescu2014variational} & Projection-based eddy viscosity variational multiscale POD \\
    2014 & \citet{san2014proper} & Smagorinsky and Chollet-Lesieur spectral vanishing eddy viscosity models  \\
    2015 & \citet{stinis2015renormalized,chorin2015discrete,li2015incorporation} &  Mori–Zwanzig  (MZ)  formalism \\
    2017 & \citet{gouasmi2017priori} & MZ ROM closures \\
    2017 & \citet{rebollo2017certified} & Reduced basis methods for the Smagorinsky closure model \\
    2017 & \citet{xie2017approximate} & Approximate deconvolution reduced order modeling \\
    2017 & \citet{benosman2017learning} & Lyapunov control theory to design learning-based closure models \\
    2018 & \citet{san2018extreme} & Extreme learning machine closure model \\
    2018 & \citet{san2017neuralrom} & Neural network closures for ROM \\
    2018 & \citet{pan2018data} & Sparse polynomial regression and neural network for closure model \\
    2019 & \citet{rahman2019dynamic} & Dynamic closure model based on a test (secondary) truncation approach \\
    2019 & \citet{stabile2019reduced} & Reduced order variational multiscale approach for turbulent flows \\
    2020 & \citet{imtiaz2020nonlinear} & \textcolor{rev}{Nonlinear closure model based on the Jacobian of the Galerkin model} \\
    2020 & \citet{reyes2020projection} & Variational multiscale ROMs\\
    2020 & \citet{xie2020closure} & Residual neural network closures \\
    2020 & \citet{wang2020recurrent} & Recurrent neural network closures \\
    2021 & \citet{mou2021reduced} & Data-driven variational multiscale ROMs \\
    2021 & \citet{gupta2020neural} & \textcolor{rev}{Neural closure models} \\
    \noalign{\smallskip}\hline
    \end{tabular}
    \label{table:closure}
\end{table*}

\subsection{Functional closure models}  \label{sec:functional}

The functional, or phenomenological, closure modeling investigations have been largely focused on the concept of eddy viscosity, which is added to the physical viscosity of the system to drain the excessive energy. This modeling concept is inspired by Kolmogorov’s ideas \cite{kolmogorov1941local,kolmogorov1991local} about the energy spectrum and energy cascade. In this section, we outline several functional (phenomenological) ROM closure modeling strategies centered around the concept of eddy viscosity.

\paragraph{\bf{Mixing-length ROM closure.}}
The first mixing-length ROM closure model was proposed by \citet{aubry1988dynamics}, who studied the wall region of the turbulent boundary layer and used a simple generalization of the Heisenberg spectral model in homogeneous turbulence to provide the eddy viscosity closure term. Specifically, the authors assumed that the deviatoric component of the Reynolds stress $\hat{\boldsymbol{\tau}}$ of the unresolved field (represented by truncated modes), acting on the resolved field (i.e., retained modes), is proportional to the strain rate $\mathbf{S}$ of the resolved field:
\begin{equation}
    \hat{\boldsymbol{\tau}} = - 2 \alpha \nu_e \mathbf{S},
\end{equation}
where $\nu_e$ is the eddy viscosity parameter and $\alpha$ is a dimensionless adjustable parameter. Moreover, they expressed the eddy viscosity term as a function of the eigenvalues and eigenfunctions of the first neglected modes based on the assumption that the energy decreases rapidly with increasing  mode index. The adequacy of this model was quantitatively validated using numerical simulations in Ref.~\cite{podvin2001adequacy}, and further investigations have been performed by Podvin and Lumley (e.g., for minimal flow unit \cite{podvin1998low} and channel flow \cite{podvin2009proper}) and also in Refs.~\cite{wang2012proper, mou2021numerical}. The two main drawbacks of the global eddy viscosity modeling approach are:
\begin{itemize}
    \item[i.] This formulation is equivalent to using Navier-Stokes equations at a lower Reynolds number. 
    \item[ii.] In this formulation, a linear closure term models nonlinear turbulence dynamics. 
\end{itemize}

\paragraph{\bf{Smagorinsky ROM closure.}}
As an improvement of the mixing-length model in Ref.~\cite{aubry1988dynamics}, the celebrated Smagorinsky model \cite{smagorinsky1963general} developed for LES has been utilized for the ROM closure problem. The eddy viscosity coefficient adapts in time in the Smagorinsky ROM closure model, but not in the mixing-length ROM closure model. Thus, the former is expected to be more accurate than the latter.
To our knowledge, the first Smagorinsky ROM closure model was proposed in~\citet{noack2002low} (see also~\cite{borggaard2008reduced}). \citet{ullmann2010pod} used the same Smagorinsky closure model of the original FOM simulations for ROMs based on LES snapshots of the turbulent flow around a circular cylinder. However, the eddy viscosity term does not appear explicitly in their ROM equation. Instead, the reconstructed velocity field is utilized to update the (spatially-varying) Smagorinsky eddy viscosity term in the FOM space, which in turn updates the corresponding model coefficients at each time step of the time integration of the ROM. \citet{borggaard2011artificial} proposed the inclusion of an artificial viscosity term in the ROM equation that resembles the one used in the Smagorinsky model even if the original FOM (for data generation) does not involve such a term. \citet{rebollo2017certified} investigated the Smagorisnky ROM closure model in a reduced basis method (RBM) setting. A rigorous numerical analysis of the Smagorinsky ROM closure model was performed in~\cite{borggaard2011artificial}, where error estimates for the time discretization were proven. To our knowledge, this represents the first numerical analysis of ROM closures. Error estimates for the time and space discretizations of the Smagorinsky ROM closure model were later proven in~\cite{rebollo2017certified} in an RBM context.

\paragraph{\bf{Dynamic SGS ROM closure.}}
The dynamic SGS model~\cite{GPMC91} is the state-of-the-art closure model in LES.
The main improvement in the dynamic SGS model over the standard Smagorinsky model is that it uses an eddy viscosity coefficient that is updated in time by using a secondary filtering operation.
The dynamic SGS closure model was extended for the first time to a ROM setting by~\citet{borggaard2008reduced} and was later investigated by~\citet{wang2012proper} in the numerical simulation of a 3D flow past a cylinder, where it yielded significantly more accurate results than both the mixing-length and the Smagorinsky ROM closure models.
A more efficient numerical discretization of the dynamic SGS ROM closure model was proposed by~\citet{rahman2019dynamic}.

\paragraph{\bf{Mode-dependent eddy viscosity closure.}}
Rather than adopting a single global eddy viscosity value $\nu_e$ for all the modes (as in the mixing-length ROM closure model), a mode-dependent eddy viscosity was proposed by \citet{rempfer1991koharente} and \citet{rempfer1993dynamics} to use a different amount of dissipation for each scale. In Refs.~\cite{rempfer1991koharente,rempfer1993dynamics}, the effective viscosity is calculated by requiring the energy variation of different modes in ROM to match the energy variation of the coherent structures in FOM. A modification to the mixing-length model can be incorporated by introducing a mode-dependent kernel. The importance of such a mode dependent kernel was first stressed by Rempfer \cite{rempfer1991koharente,rempfer1993dynamics,rempfer1994dynamics}. \citet{sirisup2004spectral} applied a vanishing viscosity kernel, which adds a small amount of mode-dependent dissipation that satisfies the entropy condition, yet retains spectral accuracy. The intrinsic stabilization scheme proposed in Ref.~\cite{kalb2007intrinsic} utilizes information from available snapshots and POD modes to define a mode-dependent stabilization. In Ref.~\cite{san2014proper}, linear, quadratic, and square-root kernels were investigated for the 1D Burgers problem.

\paragraph{\bf{Variational multiscale eddy viscosity ROM closure.}}
The variational multiscale (VMS) methods developed by Hughes and his collaborators~\cite{hughes1998variational,hughes2000large,hughes2001large} have made a significant impact in classical CFD. The VMS methods center around the principle of locality of energy transfer, which states that energy is transferred mainly between neighboring scales or modes. Since ROMs use hierarchical bases in which the large and small structures are clearly displayed, the VMS framework was naturally extended to the ROM setting. Next, we present some of the eddy viscosity ROM closure models developed in a VMS framework. \citet{borggaard2008reduced} proposed the first VMS ROM closure model, which was later investigated in~\citet{wang2012proper} in the numerical simulation of a 3D flow past a circular cylinder. The VMS ROM in Refs.~\cite{borggaard2008reduced,wang2012proper} used a \emph{three-scale} decomposition of the flow field into resolved large, resolved small, and unresolved scales, and employed the Smagorinsky model to dissipate energy only from the resolved small scales. A \emph{two-scale} decomposition of the flow field into resolved and unresolved scales was used by \citet{bergmann2009enablers} to develop a VMS ROM closure model with a residual based eddy viscosity component.
\citet{iliescu2013variational,iliescu2014variational} put forth a three-scale VMS ROM closure model, in which the ROM projection was used to construct an eddy viscosity term that acts only on the small resolved scales.

A similar three-scale VMS ROM was proposed by~\citet{roop2013proper} for a generalized Oseen problem. \citet{eroglu2017modular} developed a different three-scale VMS ROM that uses the ROM projection to add an eddy viscosity term acting only on the small resolved scales in a modular fashion. This VMS ROM was successfully tested in the numerical simulation of a turbulent channel flow at $Re_{\tau}=395$~\cite{eroglu2017modular} and was extended by~\citet{guler2019decoupled} to the Darcy-Brinkman equations with double diffusive convection. \citet{stabile2019reduced} proposed a two-scale residual-based VMS ROM closure model in which the VMS strategy is used at both the FOM and the ROM levels to ensure model consistency. \citet{reyes2020projection} (see also Ref.~\cite{reyes2020stabilized}) developed a two-scale VMS ROM which is equipped with time-dependent orthogonal subgrid scales that leverage the orthonormal nature of the POD basis. Two-scale VMS-ROMs based on orthogonal subgrid scales were used by~\citet{reyes2018reduced} for thermally coupled low Mach flows and by~\citet{tello2019fluid} for a fluid structure interaction problem. The first numerical analysis of VMS ROMs was performed in~\cite{iliescu2013variational,iliescu2014variational}, where stability and convergence were rigorously proven. Numerical analysis of VMS ROMs was also performed by~\citet{roop2013proper} and \citet{eroglu2017modular}. We also refer to the recent studies by Rubino and his coworkers\cite{rubino2020numerical,azaiez2019cure} for multi-stage ROM stabilization approaches in advection-dominated problems.

\paragraph{\bf{Finite-time thermodynamics ROM closure.}}
In the majority of the aforementioned studies, the closure term eventually appears as a linear term in the GROM (i.e., $\mathbf{\mathfrak{C}} = \mathbf{\tilde{\mathfrak{B}}} +  \mathbf{\tilde{\mathfrak{L}}} \ba$).
(The Smagorinsky and dynamic SGS ROM closure models are notable exceptions.) \citet{noack2011reduced} highlighted that energy transfer is actually caused by nonlinear mechanisms. Thus, they introduced a nonlinear eddy viscosity term $\nu_e(\ba)$ that is state-dependent. A finite-time thermodynamics (FTT) \cite{noack2008finite} approach was utilized to quantify the nonlinear eddy viscosity by matching the modal energy transfer effect as follows:
\begin{equation}
    \nu_e(\ba) = \nu_0 \sqrt{\dfrac{K(t)}{\bar{K}}},
\end{equation}
where $K(t) = \sum_{i=1}^r \dfrac{1}{2} a_i(t)^2$ represents the total turbulence kinetic energy resolved by the Galerkin expansion and $\bar{K}$ denotes its time-averaged value. This led to damping levels more consistent with energy fluctuations than those defined by a linear eddy viscosity model. The FTT-based nonlinear eddy viscosity with an energy-based scaling model was successfully applied to a 3D turbulent jet \cite{schlegel2009reduced} and a 2D mixing layer \cite{cordier2013identification}. It was further extended to a mode-dependent nonlinear eddy viscosity for a high Reynolds number flow over a square-back Ahmed body \cite{osth2014need}.

\paragraph{\bf{Efficient numerical discretization of ROM closures.}}
Although the eddy viscosity closure models discussed in this section can significantly improve the ROM accuracy, their brute-force numerical discretization can be extremely inefficient. For example, the Smagorinsky ROM closure model depends on the Frobenius norm of the deformation tensor, which is a non-polynomial nonlinearity that cannot be preassembled in the offline stage. Thus, alternative, efficient numerical discretizations have been proposed over the last decade, which we outline next. \citet{wang2011two} proposed a two-level method to avoid the brute-force discretization of the closure term onto the FOM fine mesh. Specifically, the POD bases constructed from the original fine grid resolution snapshot data were interpolated onto a coarse grid, and then they were used to efficiently compute the ROM closure term. To avoid the assembly of the FOM strain rate tensor at each time step, \citet{akhtar2012new} used the Jacobian of the GROM right-hand side as an eddy viscosity coefficient. A pre-computed eddy viscosity approach was adopted by \citet{san2015stabilized} by simplifying the nonlinear interaction in the original Smagorinsky model. \citet{san2014proper} also explored various closure approaches including constant, polynomial, and spectral vanishing viscosity models. \citet{rebollo2017certified} were the first to use an efficient hyper-reduction method~\cite{yano2019discontinuous} (i.e., EIM~\cite{barrault2004eim}) to discretize the Smagorinsky ROM closure in an RBM setting. 

\subsection{Structural closure models}  \label{sec:structural}
    
Structural closure models are generally derived through mathematical rather than phenomenological arguments. This often includes a filtering procedure, where the filtered field is assumed to have larger spatial structures than those in the original one. Therefore, the filtered flow variables require fewer modes in the ROM approximation. In other words, for the same number of modes, ROM is capable of approximating the filtered flow field more accurately than the unfiltered field. This approach is similar to LES, where the filtered flow variables can be approximated on the given coarse mesh more accurately than the original unfiltered flow variables.
In this section, we survey ROM closure models developed by using different types of ROM filtering.

\paragraph{\bf{Spatial filtering: projection.}}
Given the hierarchical nature of the ROM basis, not surprisingly, the most popular type of ROM filtering has been the ROM projection, i.e., the projection of various (nonlinear) terms living in the  $r$-dimensional ROM space spanned by the first $r$ ROM basis functions onto a smaller, $s$-dimensional ROM space spanned by the first $s$ ROM basis functions, where $s < r$.
A classical example of ROM closure models constructed by using the ROM projection is the VMS-ROMs\citep{stabile2019reduced,reyes2020projection,reyes2020stabilized,mou2021data}, which are discussed in Section~\ref{sec:functional}. The ROM projection, however, has been used to develop other types of ROM closures.
For example, the ROM projection has been utilized to construct parametrized manifolds ROM closures \citep{chekroun2014stochastic,chekroun2015approximation}, which are based on dynamical systems approaches, e.g., approximate inertial manifolds. The ROM projection has also been used to build ROM closures based on stochastic dynamical systems ideas \citep{majda2010mathematical,majda2012physics,harlim2014ensemble}. 

\paragraph{\bf{Spatial filtering: differential filter and approximate deconvolution.}}
Using the analogy between LES and ROM, we mention that a lot of ideas and techniques in image and signal processing are also applicable in ROM, and vice versa! In LES, the approximate deconvolution (AD) represents one of the most popular techniques in this class.  It is based on the deconvolution approaches developed in the image processing and inverse problems communities to recover the original signal from a blurred filtered signal.

In stark contrast to the abundance of functional closure studies (beginning in the 1980s), there are only a few structural closure models in ROM literature. The AD-ROM was proposed by \citet{xie2017approximate} for the three-dimensional flow past a circular cylinder. To construct the AD-ROM, a ROM differential filter is applied to the Navier-Stokes equations, followed by a Galerkin projection of the \emph{filtered} equations. It is usually assumed that the filtering and differentiation operators commute, and the repercussions of this assumption are investigated in Ref.~\cite{koc2019commutation}. Nonetheless, it is well known that, in general, nonlinearity and ROM spatial filtering do not commute. Therefore, the resulting equations include a \emph{filtered} nonlinear term of the \emph{unfiltered} variables (i.e., $\widehat{\mathcal{N}(\mathbf{u})}$, where the hat operator denotes the filtering process), rather than a nonlinear operator of the \emph{filtered} variables (i.e., $\mathcal{N}(\hat{\mathbf{u}}))$. A regularized deconvolution is adopted to provide the ROM approximation of the unfiltered flow variables in order to compute the nonlinear term. Thus, the filtering process increases the accuracy of the ROM in the sense that the filtered field contains larger spatial structures, and thus can be sufficiently captured by the ROM approximation. In addition, the AD technique solved the ROM closure problem by providing an estimate of the unfiltered flow variables.

\begin{remark}
We note that ROM spatial filtering has also been used to develop {\it regularized ROMs (Reg-ROMs)}, i.e., ROMs in which spatial filtering is used to smoothen (regularize) various terms in the Navier-Stokes equations and increase the numerical stability of the ROM. We emphasize that, while related, regularization and closure are different:
The latter adds a closure term, whereas the former usually does not. ROM spatial filtering has been used to develop various types of Reg-ROMs: \citet{wells2017evolve} proposed, for the first time, an evolve-then-filter approach in which the GROM (Eq.~\ref{eq:grom2}) is integrated (evolved) for one time step, after which a ROM spatial filter is applied to filter the intermediate solution obtained in the evolve step. This filtering reduces the numerical oscillations of the flow variables (i.e., adds numerical stabilization to the ROM). \citet{gunzburger2019evolve} proposed an evolve-filter-relax approach that considers the additional step of relaxation, which averages the unfiltered and filtered flow variables to control the amount of numerical dissipation introduced by the filter. Recently, Girfoglio et al.~\cite{girfoglio2021pod, girfoglio2021pressure} have investigated the evolve-filter approach in a finite volume setting. The ROM differential filter has also been used to develop the Leray Reg-ROM in Refs.~\cite{sabetghadam2012alpha,xie2018numerical,gunzburger2020leray}.
\end{remark}

\paragraph{\bf{PDF filtering: Mori--Zwanzig formalism and memory effects.}} 
A different type of filtering, based on filtering with respect to the PDF of the initial conditions, has been instrumental in adding memory effects to ROM closures, \textcolor{rev}{with rationale based on the Mori--Zwanzig (MZ) formalism \cite{mori1965transport,zwanzig1980problems,chorin2000optimal}.} More recently, the MZ formalism has been intensely used to define closures for both LES and ROM settings. Next, we outline some of these developmemnts. \citet{stinis2015renormalized} introduced a generalized MZ framework for the construction of ROMs for systems without scale separation. \textcolor{rev}{For example, assume that Eq.~\ref{eq:mz1} defines the following linear system \cite{zwanzig2001nonequilibrium,gouasmi2017priori,pan2018data,wang2020recurrent}:
\begin{equation}
    \frac{d}{dt} \begin{bmatrix} \ba \\ \boldsymbol{b} \end{bmatrix} =  \begin{bmatrix} A_{11} & A_{12} \\ A_{21} & A_{22} \end{bmatrix} \begin{bmatrix} \ba \\ \boldsymbol{b} \end{bmatrix}. \label{eq:mz3}
\end{equation}
}
\textcolor{rev}{The evolution of the unresolved state $\boldsymbol{b}$ can be evaluated as follows (assuming that $\ba$ is known) \cite{zwanzig2001nonequilibrium,gouasmi2017priori,wang2020recurrent}:
\begin{equation}
    \boldsymbol{b}(t) = \int_{0}^{t} e^{A_{22}(t-s)}A_{21} \ba(s) ds + e^{A_{22}t} \boldsymbol{b}(0). \label{eq:mzb}
\end{equation}
Therefore, Eq.~\ref{eq:mz2} for the dynamics of $\ba$ can be written as
\begin{align}
    \frac{d\ba}{dt} &= A_{11} \ba +  A_{12}\boldsymbol{b} \nonumber\\
    &= A_{11} \ba \nonumber \\
    &+ A_{12} \int_{0}^{t} e^{A_{22}(t-s)}A_{21} \ba(s) ds \nonumber \\
    &+ A_{12} e^{A_{22}t} \boldsymbol{b}(0). \label{eq:mz4}
\end{align}
Eq.~\ref{eq:mz4} expresses the dynamics of the resolved scales using a Markovian term (i.e., $A_{11} \ba$, which depends only on the current value of $\ba$), a memory integral term depending on the history of the resolved scales $\ba$, and a term describing the contribution of the initial conditions. This derivation can be extended to nonlinear settings where, for instance, the nonlinear ordinary differential equation can be written as a linear partial differential equation using the Liouville operator. The exact evolution equations for the reduced state can be written as
\begin{align}
    \frac{d\ba}{dt} &= \bff_{\ba}(\ba,0) \nonumber \\ 
    &+ \int_{0}^{t} \mathcal{K}(\ba(s),t-s) ds + \mathcal{O}(\ba(0),\boldsymbol{b}(0)), \label{eq:mz5} \\
    &\approx \bff(\ba) + \text{Closure}(\ba) . 
\end{align}
In Eq.~\ref{eq:mz5}, $\mathcal{K}$ is called the memory kernel and $\mathcal{O}$ designates the contribution from the initial conditions. The memory integral term implies that the accurate resolution of $\ba$ comprises a non-Markovian contribution. However, the direct computation of Eq.~\ref{eq:mz5} is generally prohibitive, and estimation of the memory-effect is often sought.} 

\citet{li2015incorporation} included a great discussion on incorporation of memory effects in coarse-grained modeling via the MZ formalism. A discrete approach to stochastic parametrization, dimension reduction, and their connections to the MZ formalism of statistical physics has been proposed by \citet{chorin2015discrete}. In an LES setting, \citet{parish2017dynamic} framed the MZ closure modeling approach by exploiting similarities between two levels of coarse-graining via the Germano identity of fluid mechanics and by assuming that memory effects have a finite temporal support. The concept has been also generalized to provide a mathematically consistent framework for the construction of ROMs of dynamical systems \cite{gouasmi2017priori}. Moreover, \citet{parish2017unified} established an analogy between MZ and VMS approaches.

\paragraph{\bf{Ensemble averaging.}}
\citet{noack2011reduced} first used ensemble averaging to construct a finite-time thermodynamics (FTT) \cite{noack2008finite} framework. \textcolor{rev}{\citet{gunzburger2017ensemble} built ensemble-based POD ROMs, where the nonlinear advection term in the Navier-Stokes equations is replaced by a linear term in the equations for the resolved scales. This \emph{linearization} is performed by using an ensemble of solution trajectories by propagating an ensemble of ROMs with varying parameters and/or initial conditions and updating the ensemble average at each time step. Later on, this ensemble-based approach was equipped with Leray regularization to develop regularized ROMs for high values of Reynolds number \citep{gunzburger2020leray}}. 


\paragraph{\bf{Time averaging.}}
\citet{selten1995efficient,selten1997statistical} used time averaging to develop ROM closures. \textcolor{rev}{In particular, by estimating the rate at which the ROM trajectory drifts away from the projection of the FOM solution on the ROM subspace, \citet{selten1995efficient} added a linear damping to expand the doubling-time of the error resulting from the modal truncation}. \citet{berselli2020long} developed mathematical support for eddy viscosity modeling of time-averaged ROM closures. While being interested in a statistical equilibrium problem exploring possible forward and backward average transfer of energy among ROM basis functions, they found that the time-averaged energy exchange from low index POD modes to high index POD modes is positive for long enough time intervals. \textcolor{rev}{This study provides, for the first time, mathematical support for the ROM eddy viscosity methodology, where the energy transfer to the truncated modes is modeled by employing extra viscous dissipation.}


\paragraph{\bf{Calibrating the POD space with a Navier-Stokes based side constraint.}}
The last modeling approach that we discuss in this section is that proposed by \citet{balajewicz2013low}. Although this approach does not add a ROM closure model, it does leverage mathematical arguments to model the effect of the truncated modes. In this approach, the POD subspace is subjected to a Navier-Stokes based side constraint. Specifically, the power balance for the fluctuation energy is required to be satisfied by the attractor data after the Galerkin projection on the adjusted POD space. This procedure can be conceptualized as rotating the POD subspace into a more dissipative regime, in which the extra dissipation is now performed by more dissipative POD modes.

\subsection{Stochastic closure models} \label{sec:sde}

Although we are mainly focusing on deterministic closure modeling in this review, we emphasize that the need for stochastic modeling was already formulated in \citet{aubry1988dynamics} to avoid statistically nonstationary behaviour for some homoclinic orbits. \textcolor{rev}{The dynamics of the unresolved scales, and hence their interactions with the resolved scales are unknown. Thus, we can only form an approximate idea of how the truncated modes behave and affect the ROM solution. Even with the best closure model, we can never be certain about its accuracy in practical settings. Thus, it is natural to model the dynamics of the unresolved modes using a random or \emph{stochastic} process, from which we can infer the unresolved modes' contribution to the evolution of large scales in a statistical sense.} We refer to \citet{leith1996stochastic,chorin2009stochastic,majda2010mathematical,majda2012physics,harlim2014ensemble,majda2015statistical,resseguier2015stochastic, chorin2015discrete,lu2017data,lu2020data,sieber2021stochastic,chekroun2014stochastic} for detailed discussions on the probabilistic modeling of such random/chaotic systems as well as the development of statistically accurate ROMs and stochastic closure models \cite{wilks2005effects,sapsis2009dynamically,sapsis2013attractor,sapsis2013statistically,sapsis2013statistically2,arnold2013stochastic,pulido2018stochastic,chattopadhyay2020data,gagne2020machine}. We also note that nonparametric stochastic modeling approaches have been proposed for representative stochastic It\^{o} drift diffusion forecast models \cite{berry2015nonparametric}. Next, we briefly outline a few of these strategies.

\textcolor{rev}{Stochastic closure approaches seek to account for the effects of the unresolved scales on the long-term statistics of the resolved scales. In particular, the closure term is modeled by a stochastic process, usually represented by Markovian and/or non-Markovian dynamics with a random forcing (e.g., random noise). For a truncated ROM of the Kuramoto-Sivashinsky system, \citet{lu2017data} defined a discrete-time closure term $z^n$ at time $t_n$ as follows:
\begin{equation}
    z^n = \Phi^n + \xi^n,
\end{equation}
where $\xi$ is a sequence of independent identically distributed random variables, which are sampled from Gaussian distributions and characterize the stochastic component of the closure, while $\Phi$ is a function of current and past values of the resolved scales $\ba$ and the forcing $\xi$. The authors used the nonlinear autoregression moving average with exogenous input (NARMAX) approach to parameterize $\Phi$. A similar approach was adopted in Ref.\cite{chorin2015discrete} for the Lorenz 96 model and Ref.\cite{lu2020data} for the stochastic Burgers equation. The multiscale Lorenz 96 model \cite{lorenz1996predictability}, which has been considered as a non-trivial test problem for stochastic paramerization in geophysical fluid dynamics studies, can be written as
\begin{align}
    \frac{d X_i}{dt} &= -X_{i-1} (X_{i-2} - X_{i+1}) - X_i - \frac{hc}{b} \sum_{j=1}^J Y_{j,i} + F, \label{eq:l96slow}\\
    \frac{d Y_{j,i}}{dt} &= -cbY_{j+1,i} (Y_{j+2,i} - Y_{j-1,i}) - c Y_{j,i} + \frac{hc}{b} X_{i} \label{eq:l96fast},
\end{align}
where Eq.~\ref{eq:l96slow} represents the evolution of slow, high-amplitude variables $X_i~(i=1,\dots,I)$, and Eq.~\ref{eq:l96fast} describes the evolution of coupled fast, low-amplitude variables $Y_{j,i}~(j=1,\dots,J)$. In order to investigate different closure approaches, $X$ can be considered as the resolved scales, while $Y$ can be considered as the unresolved ones. Therefore, Eq.~\ref{eq:l96fast} is assumed to be unknown and is only used for generating \emph{true} data. Furthermore, the form of the term $- \frac{hc}{b} \sum_{j=1}^J Y_{j,i}$, representing the contribution of $Y$ to the dynamics of $X$, is also assumed to be unknown. A closure term is parameterized as a function of the resolved scales. Although the Lorenz 96 equations are determinstic, \citet{wilks2005effects} showed the existence of multiple closure values that are consistent with any given large-scale variable, i.e., that different values of the closure term yield \emph{statistically} similar results. Therefore, \citet{wilks2005effects} defined the closure term using both deterministic and stochastic components. Specifically, the author used a fourth-order polynomial fitting for the deterministic part that represents the average trend, and a first-order auto-regression model for the stochastic part that defines the deviation of different realizations from the fitted curve. \citet{arnold2013stochastic} explored several parametrization schemes for the stochastic component, including additive and multiplicative noise.}

\textcolor{rev}{Although the distinction between resolved and unresolved variables in the multiscale Lorenz 96 system is not driven by a Galerkin truncation as is the case for most projection-based ROMs (which is the focus of the current review), the same arguments apply in both scenarios. For example, \citet{memin2014fluid} assumed that the flow field is decomposed into a deterministic resolved component and a generalized random field that models the unresolved flow component and all the uncertainties in the flow. In other words, the projection of the flow field onto the truncated space is treated as a \emph{realization} or sample of the stochastic component of the flow and is modeled using Brownian motion. \citet{resseguier2015stochastic} provided numerical investigations of this methodology using POD-Galerkin projection for flow past cylinder. Nonetheless, we believe that this is an open research area that needs fresh ideas to translate statistical closure strategies \cite{zhou2021turbulence} from turbulence modeling to the ROM arena.}


\section{Data-driven closure modeling} \label{sec:data}

With the abundant supply of big data, open-source cutting edge and easy-to-use machine learning libraries, cheap computational infrastructure, and high quality, readily available training resources, data-driven closure modeling has become very popular. Since projection-based ROMs are usually constructed from snapshot data (either collected experimentally or computationally), it is natural to further exploit this set of data to estimate the closure term efficiently. 

In this section, we survey data-driven closure modeling approaches in which the closure problem is cast into a regression task.
The majority of the recent data-driven closure studies can be viewed as a regression task, where the closure model is defined partially or completely as a function of available information (e.g., resolved scales). 
We first discuss the early investigations based on classical least-squares approaches. Then, we introduce ML tools that perform this regression task using neural networks and Gaussian processes regression. Finally, we explore  legacy data assimilation, parameter estimation, and system identification tools that can efficiently be used to solve the closure problem. We note that some subsections are entirely devoted to closure modeling (Sections~\ref{sec:trajectory-regression-vs-model-regression}, \ref{sec:least-squares-regression}), some only partially address closure modeling (Sections~\ref{sec:neural-networks-regression}, \ref{sec:da}), and some do not address closure modeling at all (Sections~\ref{sec:kernel-regression}, \ref{sec:oi}, \ref{sec:id}).
Although the approaches in the latter subsections have not yet been used for closure modeling, we believe that they will soon make an impact in this dynamic research field.

\subsection{Trajectory regression vs. model regression} \label{sec:trajectory-regression-vs-model-regression}

There are two main schools of thought in data-driven ROM closure modeling: (i) trajectory regression, and (ii) model regression.

The {\it trajectory regression} approach aims at finding the ROM closure model $\mathbf{\mathfrak{C}}$ that yields the best ROM trajectory.
In this approach, the following constrained regression problem is solved:
\begin{eqnarray}
    \begin{aligned}
        & \underset{\mathbf{\mathfrak{C}} \text{ parameters}}{\text{minimize}} & & \| \ba^{\small\rm ROM} - \ba^{\small\rm FOM} \|^2 , \\
        & \text{subject to}             & & \ba^{\small\rm ROM} \text{~solves closed ROM (Eq.~\ref{eq:gromC})}
    \end{aligned}
    \label{eqn:trajectory-regression}
\end{eqnarray}
where $\ba^{\small\rm FOM}$ is the vector of ROM coefficients computed with the FOM data and $\ba^{\small\rm ROM}$ is the vector of ROM coefficients yielded by the closed ROM.

The {\it model regression} approach aims at finding the ROM closure model $\mathbf{\mathfrak{C}}$ that yields the best ROM closure model.
In this approach, the following unconstrained regression problem is solved:
\begin{eqnarray}
    \begin{aligned}
        & \underset{\textbf{c}}
        {\text{minimize}} & & \| \mathbf{\mathfrak{C}}^{\small\rm ROM} - \mathbf{\mathfrak{C}}^{\small\rm FOM} \|^2 ,    \end{aligned}
    \label{eqn:model-regression}
\end{eqnarray}
where $\mathbf{\mathfrak{C}}^{\small\rm FOM}$ is the ROM closure model computed with the FOM data, $\mathbf{\mathfrak{C}}^{\small\rm ROM}$ is the postulated ROM closure model form, and $\textbf{c}$ is the vector of parameters used to define the ROM closure model form. We note that the model regression approach is similar in spirit to the {\it a priori} testing used in LES~\cite{sagaut2006large}.

We emphasize that the two approaches are fundamentally different. The trajectory regression is a black-box approach in which the precise formula for the ROM closure term, $\mathbf{\mathfrak{C}}$, is not actually known. Instead, the trajectory regression first  postulates a model form for $\mathbf{\mathfrak{C}}$ (e.g., by using one of the functional models in Section~\ref{sec:functional}), and then solves the constrained optimization problem Eq.~\ref{eqn:trajectory-regression} to find the closure model parameters that yield the most accurate ROM trajectory (i.e., the trajectory $\ba^{\small\rm ROM}$ that is closest to the projection of the FOM data on the ROM basis, $\ba^{\small\rm FOM}$). In contrast, the model regression first employs one of the filters described in Section~\ref{sec:structural} to determine a precise formula for the ROM closure term, $\mathbf{\mathfrak{C}}^{\small\rm FOM}$. Then, it postulates a model form for $\mathbf{\mathfrak{C}}^{\small\rm ROM}$. Finally, it solves the unconstrained optimization problem Eq.~\ref{eqn:model-regression} to find the closure model parameters that yield the most accurate ROM closure model (i.e., the closure model $\mathbf{\mathfrak{C}}^{\small\rm ROM}$ that is closest to the ``true'' closure model, $\mathbf{\mathfrak{C}}^{\small\rm FOM}$, computed from FOM data).

There are pros and cons for both approaches. The trajectory regression is conceptually simpler than the model regression since it does not need to determine the actual form of the ROM closure term. The trajectory regression is also more flexible than the model regression since it can model not only the ROM closure term, but also other sources of ROM uncertainty, such as the numerical discretization error and the missing data. Finally, according to Noack's conjecture~\cite{noack2005need}, the model regression is more accurate in the predictive regime (i.e., outside the training interval), whereas the trajectory regression is more accurate in the reconstructive regime (i.e., inside the training interval). Noack motivated his conjecture by noting that using data to match models appears more robust to perturbations than using data to match trajectories. To our knowledge, Noack's conjecture has not been investigated numerically.

\subsection{Least-squares regression}   \label{sec:least-squares-regression}

In this section, we survey the data-driven ROM closure models that use a least-squares formulation in the optimization problems given by Eq.~\ref{eqn:trajectory-regression} and Eq.~\ref{eqn:model-regression}.

\paragraph{Trajectory regression.}

Given its conceptual simplicity, the trajectory regression has been used from the earliest days of reduced order modeling to develop closures.
The general idea used to develop these closure models is simple:
(i) Postulate a ROM closure model form, either functional (such as the eddy viscosity models surveyed in Section~\ref{sec:functional}) or structural (such as the models surveyed in Section~\ref{sec:structural}). 
(ii) Use a least-squares problem in Eq.~\ref{eqn:trajectory-regression} to determine the various parameters in the postulated model form.

\emph{Functional models}: \ 
Probably the first functional trajectory regression closure is the mixing-length model proposed in the pioneering work by \citet{aubry1988dynamics}, in which trajectory regression is used to determine the mixing-length constant (see~\cite{wang2012proper} for related work).
A least-squares trajectory regression was also used by~\citet{wang2012proper} to determine the eddy viscosity constants in the Smagorinsky and VMS closure models.
Further improvements to the least-squares trajectory regression of eddy viscosity ROM closure models were proposed by~\citet{osth2014need} and \citet{protas2015optimal}.
The eddy viscosity trajectory regression has also been used by \citet{stabile2019reduced,reyes2020projection}, and \citet{bergmann2009enablers}.

\emph{Structural models}: \ 
Probably the first structural trajectory regression closures are those constructed with \emph{calibration} methods, which have been introduced to directly modify (or calibrate) the GROM polynomial coefficients (i.e., $\mathbf{{\mathfrak{B}}}$, $\mathbf{{\mathfrak{L}}}$, and $\mathbf{{\mathfrak{N}}}$), rather than introducing an additional closure term. \citet{buffoni2006low} calibrated the constant and linear terms (i.e., $\mathbf{{\mathfrak{B}}}$ and $\mathbf{{\mathfrak{L}}}$), while leaving the nonlinear term, $\mathbf{{\mathfrak{N}}}$, as derived from the Galerkin projection step. The modified coefficients are then found using a pseudo-spectral method such that the model prediction is as close as possible to the actual reference solution \cite{galletti2007accurate}. An extension to calibrate \emph{all} the polynomial coefficients (linear and quadratic) was employed by \citet{couplet2005calibrated}, where the cost function is defined to penalize the deviation of the calibrated ROM behavior with respect to the projection of true snapshot data. Moreover, \citet{perret2006polynomial} considered a cubic polynomial to represent the dynamics of the POD modal amplitudes for supersonic jet-mixing layer data, and adopted a least-squares regression to define the coefficients of the polynomial. 
\citet{baiges2015reduced} used a calibration method in a VMS framework to build a structural trajectory regression closure. Specifically, the authors postulated a linear model for the unresolved sub-scale term as a function of the resolved field (i.e., $\mathbf{\mathfrak{C}} = \mathbf{\tilde{\mathfrak{B}}} +  \mathbf{\tilde{\mathfrak{L}}} \ba$), and then solved the constrained least-square problem Eq.~\ref{eqn:trajectory-regression} for the components of $\mathbf{\tilde{\mathfrak{B}}}$ and $\mathbf{\tilde{\mathfrak{L}}}$.

An assessment of various calibration techniques using the two-dimensional ﬂow around a cylinder can be found in Ref.~\cite{cordier2010calibration}. A vital merit of the calibration methods is that the computational costs of these methods are reasonable since they employ the temporal part of the POD information for the regression task, while the Galerkin projection method exploits the much more voluminous spatial POD information to construct the ROM polynomial. 


\paragraph{Model regression.}

These closure models are constructed as follows:
(i) Use the filters surveyed in Section~\ref{sec:structural} to determine a mathematical formula for the ``true'' closure model, $\mathbf{\mathfrak{C}}^{\small\rm FOM}$, computed from FOM data.
(ii) Postulate a ROM closure model form for the ROM closure model, $\mathbf{\mathfrak{C}}^{\small\rm ROM}$. 
(iii) Use the least-squares problem in Eq.~\ref{eqn:model-regression} to determine the various parameters in the postulated model form.
To our knowledge, the vast majority of ROM closure model forms that have been proposed in this direction are of structural type. 
A notable exception is the model proposed by~\citet{hijazi2020data}, which uses finite volume RANS data in conjunction with a model regression approach to determine the eddy viscosity component of the ROM.

A model regression closure that uses the ROM projection as a spatial filter is the data-driven VMS-ROM model proposed by~\citet{mou2021data}.
A two-scale version was investigated in \cite{xie2018data,mou2020data}, and a three-scale version was proposed in~\cite{mou2021data}.
Linear~\cite{mou2020data}, quadratic~\cite{xie2018data,mou2021data}, and even cubic~\cite{mou2018cross} terms were used for the model form:
$
\mathbf{\mathfrak{C}} = \mathbf{\tilde{\mathfrak{B}}} 
+ \mathbf{\tilde{\mathfrak{L}}} \ba 
+ \ba^\top \mathbf{\tilde{\mathfrak{N}}} \ba 
+ \ba^\top ( \ba^\top \mathbf{\tilde{\mathfrak{N}}} \ba ) 
$. 
A data-driven VMS-ROM to increase the  pressure accuracy was proposed in \cite{ivagnes2021data}. The verifiability of the data-driven VMS-ROM was proven by Koc et al~\cite{koc2021verifiability}. Other model regression closures that use the ROM projection as a spatial filter were employed to build parameterized manifold closures by Liu and his collaborators~\cite{chekroun2014stochastic} 
, and by Lu and his coworkers~\cite{lu2020data,lin2019data,lu2017data} to construct stochastic ROM closures. A model regression closure that uses the ROM differential filter was proposed by~\citet{koc2019commutation}. The resulting data-driven LES-ROM uses the ROM differential filter to determine a mathematical formula for the ``true'' closure model, $\mathbf{\mathfrak{C}}^{\small\rm FOM}$, computed from FOM data. To our knowledge, the data-driven LES-ROM in~\citet{koc2019commutation} is the only model regression approach that utilizes a {\it spatial filter} (i.e., the differential filter) instead of the commonly used ROM projection. 

A model regression closure that uses the PDF filtering in an MZ setting was proposed by Duraisamy, Parish, and their collaborators~\cite{gouasmi2017priori,parish2018residual,parish2020adjoint}. In the MZ framework, PDF filtering is used to express the ``true'' closure model, $\mathbf{\mathfrak{C}}^{\small\rm FOM}$, as a memory term, which is then approximated by using the FOM data and solving the least-squares problem Eq.~\ref{eqn:model-regression}. Moreover, a model regression closure that uses time filtering was proposed by~\citet{selten1997statistical} and utilized in the numerical simulation of a barotropic model.


\citet{mohebujjaman2019physically} used physical constraints in the model regression closure to improve the stability and accuracy of the data-driven VMS-ROM model~\cite{mou2021data}. Specifically, they equipped the least-squares problem Eq.~\ref{eqn:model-regression} with physical constraints to enforce the regressed matrix and tensor to have similar characteristics as the GROM operators (e.g., $\mathbf{\tilde{\mathfrak{L}}}$ being negative semi-definite and $\mathbf{\tilde{\mathfrak{N}}}$ being energy conserving). 
A similar quadratic formula was adopted in Ref.~\cite{pawar2020evolve} to recover the hidden physical processes (e.g., source terms) for system with incomplete governing equations.



\subsection{Neural network regression} \label{sec:neural-networks-regression}
\textcolor{rev}{The introduction of neural network regression into ROM was highly motivated by the desire to construct purely data-driven nonintrusive ROM (NIROM) frameworks \cite{yu2019non,fresca2021comprehensive}, which solely rely on data to learn the dynamics of the relevant solution manifold without the need to access the governing equations. Non-intrusive approaches are attractive due to their portability since they do not necessarily require the exact form of the equations or the methods used to generate the data. In addition, non-intrusive models offer a unique advantage in multidisciplinary collaborative environments, where only data can be shared without revealing the proprietary or sensitive information. Non-intrusive approaches are also useful when the detailed governing equations of the problem are unknown. This modeling approach can benefit from the enormous amount of data collected from experiments, sensor measurements, and large-scale simulations to build NIROMs based on the assumption that data is a manifestation of \emph{all} the underlying dynamics and processes.} 

\begin{figure}[ht]
\centering
\includegraphics[width=0.9\linewidth]{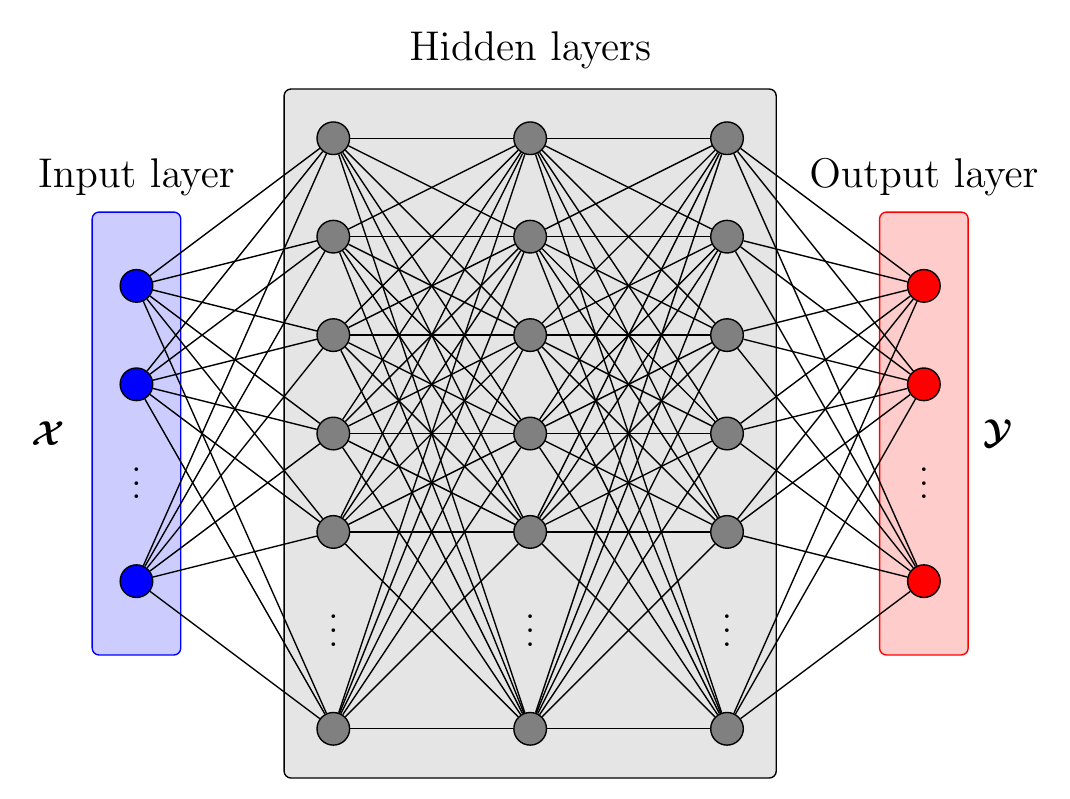}
\caption{A schematic diagram of a typical feed-forward neural network with an input layer, hidden layers, and an output layer. \textcolor{rev}{We note that these general-purpose dense deep network architectures have been evolving to more specific neural network designs \cite{goodfellow2016deep}. We discuss some of them briefly in Section~\ref{sec:pgml}.}}
\label{fig:dnn}
\end{figure}

Machine learning tools, in particular artificial neural networks (ANNs) equipped with the universal approximation theorem \cite{csaji2001approximation}, have been widely exploited in this regard. A typical feed-forward neural network is depicted in Figure~\ref{fig:dnn}, where a mapping $\boldsymbol{\mathcal{M}}$ from the input $\boldsymbol{\mathcal{X}}$ to the output $\boldsymbol{\mathcal{Y}}=\boldsymbol{\mathcal{M}}(\boldsymbol{\mathcal{X}})$ is inferred through a learning algorithm. For transient flows, a single-layer feed-forward neural network was proposed by \citet{san2019artificial} to provide accurate predictions of the ROM coefficients with varying control parameter values, using sequential and residual approaches. In the sequential approach, a mapping from the current values of $\ba$ to their future values is approximated. Moreover, the input layer is augmented with the acting Reynolds number and the time. That is,  $\boldsymbol{\mathcal{X}}= \{\text{Re},t_n, \ba(t_n)\}$, while $\boldsymbol{\mathcal{Y}}= \{\ba(t_{n+1})\}$. On the other hand, the residual implementation relies on learning the deviation of the future state from the current values (i.e., $\boldsymbol{\mathcal{Y}}= \{\ba(t_{n+1})-\ba(t_{n})\}$). \citet{pawar2019deep} employed deep neural networks (DNN) to bypass the Galerkin projection step and build a fully NIROM for the two-dimensional Boussinesq equations with a differentially heated cavity flow setup at various Rayleigh numbers. In particular, the evolution of the POD modal amplitudes $\ba(t_{n+1})$ was predicted from their past values using residual and backward difference scheme formulas. The application of variants of ANNs as regression models for the dynamics of low-order states (e.g., POD amplitudes) has gained substantial popularity \cite{hesthaven2018non,wang2019non,wu2020data}, owing to the availability of open-source and user-friendly ML libraries. This is a hot topic and dozens of new papers appear every week in different journals and conferences all over the world, dealing with different aspects of NIROM based on ANNs (e.g., different architectures, test bed problem, and error bounds).

Recurrent neural networks (RNNs) are very effective for sequence predictions in numerous applications, e.g., speech recognition and translation. RNNs contain loops that allow them to retain information from one time step to another so as to enforce the temporal dependencies. A deep residual recurrent neural network was utilized by \citet{kani2017dr} for the model reduction of nonlinear dynamical systems. However, one of the limitations of RNNs is vanishing (or exploding) gradient to capture the long-term dependencies, stemming from the repetitive multiplication of gradient with potentially ill-conditioned weight matrices during the back-propagation learning algorithm. The long short-term memory (LSTM) neural networks mitigate this issue by employing a gating mechanism that allows information to be forgotten. \citet{vlachas2018data} trained an LSTM to predict the derivative of $\ba$ with respect to time from a short history of $\ba$ values, where a first-order forward difference scheme was adopted to represent the temporal derivative. They also combined the LSTM with a mean stochastic model to cope with attractor regions that are not captured in the training set. \citet{mohan2018deep} explored the bidirectional variant of LSTM, employing two LSTM networks: one in the forward and the other in the reverse direction, for NIROM of forced isotropic turbulence and magneto-hydrodynamic turbulence using the Johns Hopkins turbulence database \cite{graham2016web}. \citet{rahman2019nonintrusive} utilized LSTMs for the NIROM implementation of the two-dimensional single-layer quasi-geostrophic ocean circulation model. A sliding window approach was adopted to predict the evolution of the POD amplitudes. In another interesting article, 
\citet{wang2020recurrent} utilized a conditioned LSTM for the memory term in the GROM equations, representing the closure model, for parametric systems.

Instead of using ML to entirely replace the GROM with NIROM, data-driven ML can be utilized along with the physics-based GROM to construct the closure model. This hybrid approach was adopted in Ref.~\cite{san2018extreme}, which used a dissipative term employing an eddy viscosity coefficient and utilized a single layer extreme learning machine (ELM) to estimate a modal $\nu_e$ as a function of the mode index, GROM right-hand side (RHS), and modal amplitudes. Furthermore, a clipping procedure was carried out by discarding negative values of $\nu_e$. ML tools can be exploited to provide the numerical value of the closure term, without constraints on the functional form of the closure model. \citet{san2018neural,san2018machine} utilized an ELM \cite{huang2006extreme} to learn the value of the closure term as a function of the GROM RHS, i.e., $\mathbf{\mathfrak{C}}= f(\mathbf{\mathfrak{B}} +  \mathbf{\mathfrak{L}} \ba +  \ba^\top \mathbf{\mathfrak{N}} \ba)$. For training purposes, the true closure term is computed from the projection of the pure evolution of the PDE onto the reduced space. In other words, Eq.~\ref{eq:uode} is first evaluated in the FOM space, then projected onto the basis functions, while the GROM RHS (Eq.~\ref{eq:grom2}) is computed directly in the reduced space.

\citet{wan2018data} utilized LSTM architectures to learn the mismatch resulting from the imperfect GROM RHS as a function of the sequence of past values of the resolved state. The ML model is exploited to assist the imperfect model whenever data is available, while for locations with sparse data, the GROM still provides an acceptable baseline for the prediction of the system state. The estimation of the closure term as a function of the time history of the resolved scales has roots in the Mori-Zwanzig formalism \cite{chorin2002optimal,chorin2009stochastic} and the memory embedding of LSTM implementation for closure modeling is supported by the Takens embedding theorem \cite{takens1981detecting}. \textcolor{rev}{\citet{gupta2020neural} employed RNNs to learn the non-Markovian term appearing in Eq.~\ref{eq:mz5} using neural delay differential equations with discrete-delays as follows:
\begin{equation}
     \frac{d\ba}{dt} \approx \bff(\ba) + g_{RNN}(\ba(t), \ba(t-\tau_1), \dots, \ba(t-\tau_K);\theta),
\end{equation}
where $K$ is the number of discrete-delays and $\theta$ represents the neural network weights. For distributed-delays, the ROM dynamics is written as
\begin{equation}
     \frac{d\ba}{dt} \approx \bff(\ba) + g\bigg(\ba(t), \int_{t-\tau_2}^{t-\tau_1} h(\ba(\tau),\tau) d\tau,t\bigg),
\end{equation}
where the delay is distributed over past time periods $t-\tau_2$ and $t-\tau_1$. \citet{gupta2020neural} approximated the $g$ and $h$ functions defining the delay term using two different coupled neural networks as follows:
\begin{align}
     \frac{d\ba}{dt} &\approx \bff(\ba) + g_{NN}(\ba(t), y(t), t ; \theta_g) \\
     \frac{dy}{dt} &\approx h_{NN}(\ba(t-\tau_1), t-\tau_1 ; \theta_h) - h_{NN}(\ba(t-\tau_2), t-\tau_2 ; \theta_h),
\end{align}
where the memory effect can be embedded without the use of any specific recurrent neural network. The authors showed that the non-Marovian closure outperforms its Marovian counterpart (with no time delays).}


\subsection{Kernel regression} \label{sec:kernel-regression}
Gaussian process regression (GPR) \cite{rasmussen2003gaussian} has the advantage of the simultaneous prediction of system's dynamics and the associated uncertainty. \textcolor{rev}{Although GPR provides a powerful tool for probabilistic inference that enables modelers to strike a balance between model complexity and data fitting \cite{raissi2018hidden}, its use is often limited to relatively small training data sets due to the well-known cubic scaling characteristics of the Gaussian processes. However, fast algorithms using approximate matrix-vector products can be utilized for large data sets \cite{raykar2010fast,chalupka2013framework}.}
GPR has gained prominence providing surrogate models for complex and multidimensional systems \cite{forrester2007multi,perdikaris2015multi,feldstein2020multifidelity}. In ROM applications, these confidence measures can be particularly informative when the ROM dimension is lower than the intrinsic dimension of the system. \citet{wan2017reduced} formulated a stochastic model based on GPR dynamics and utilized a Monte-Carlo framework for the forecast of the system's state and corresponding uncertainty. \citet{maulik2020latent} utilized a Gaussian process emulator for the dynamical evolution of the latent space state variables, obtained from POD and autoencoder compression, for the shallow water equations. \textcolor{rev}{We also note that \citet{raissi2018hidden} developed a Gaussian process framework to learn PDEs from relatively small quantities of data.} 
\citet{xiao2015non} applied a second order Taylor series scheme and a Smolyak sparse grid collocation method to calculate the POD modal coefficients at each time step from their values at earlier time steps. A radial basis function (RBF) multi‐dimensional interpolation was used in Ref.~\cite{xiao2015non2} for similar purposes. RBF interpolations have been also utilized for parameterized problems as mappings from the parameter space to the ROM space \cite{rajaram2020randomized}.

\subsection{Data assimilation and error correction} \label{sec:da}

An excellent discussion of the similarities between ML and data assimilation (DA) tools has been recently provided by Alan Geer \citep{geer2021learning}. \textcolor{rev}{The synergistic integration of ML and DA is essential for developing improved approaches \cite{tang2001coupling,liaqat2003applying,brajard2020combining,brajard2020combiningb,bonavita2020machine,farchi2020using,bocquet2020bayesian,farchi2021comparison,pawar2021data,pawar2021nonintrusive}.} Looking at the ROM framework from a data assimilation point of view opens up innovative avenues to tackle the closure problem. Ideas from optimal control theory, data assimilation, and parameter estimation were proven to be valuable in this regard. Data assimilation is a generic framework combining the available observations with the underlying dynamical principles governing the system to estimate the physical quantities of interest. This is usually accomplished by starting from a background solution and computing an optimal estimation of the true state of the system that minimizes the discrepancy between model predictions and collected observations. This minimization problem is solved while taking into account the respective statistical confidence of different observations, background solution, and model uncertainty \cite{ahmed2020pyda}. In order to emphasize the model's error (due to truncation), \citet{d2007variational} added a Gaussian variable to the ROM equation. This is similar to the weak variational data assimilation framework implemented in Ref.~\cite{artana2012strong} for ROMs using real experimental conditions with noisy particle image velocimetry data. The GROM is augmented with an additive stochastic control variable, representing the model's uncertainty that reflects the effect of unresolved scales on the resolved dynamics. Estimating such uncertainty function can be incorporated to improve the ROM predictions. \citet{zerfas2019continuous} utilized the nudging algorithm to improve predictions by adding a feedback control term that nudges the ROM approximation towards the reference solution corresponding to  the observed data. The authors also presented a strategy to dynamically adjust the nudging parameter by controlling the dissipation arising from the nudging term, as well as a numerical analysis of the proposed DA-ROM. A combination of the nudging methodology and LSTM framework was adopted by~\citet{ahmed2020reduced,ahmed2020nudged} to correct the ROM trajectory, considering the initial condition mismatch and GROM model deficiency.

DA has been also utilized to \emph{calibrate} the ROM coefficients so that the ROM predictions agree with available observations. \citet{cordier2013identification} adopted a four-dimensional variational (4DVAR) formulation to tune the computed ROM coefficients, where the background values were obtained from standard Galerkin projection. The projection of snapshot data onto the POD basis was treated as synthetic observations of the reduced system's state. Although a good match was observed \emph{within} the assimilation window, the model's stability on the forecast window was not ensured. This is similar to the strong variational formulation in Ref.~\cite{d2007variational} aiming at directly correcting the model's coefficients assuming a deterministic dynamical model. 

DA tools can also be exploited to provide a good estimate of the free parameters in classical closure models. For example, \citet{cordier2013identification} adopted the nonlinear eddy viscosity model from \cite{noack2008finite,schlegel2009reduced} and applied the 4DVAR framework to estimate the eddy viscosity parameter to increase the physical reliability of the model \emph{beyond} the assimilation window. More recently, \citet{ahmed2020forward} used a linear eddy viscosity model and exploited the forward sensitivity method (FSM) to compute and update the mode-dependent eddy viscosity parameters. Given the plummeting costs of sensors and the potential of ROM in real time monitoring and control, we emphasize that DA appears to be a good candidate for future developments leveraging the increasingly available heterogeneous measurement data to build more robust ROM closures. 

\subsection{Operator inference approaches} \label{sec:oi}

An \emph{operator inference} (OI) approach was proposed by \citet{peherstorfer2016data} 
to infer the ROM operators from data. Next, we briefly outline the OI approach. First, we note that the quadratic term in Eq.~\ref{eq:grom2} can be written as $[\ba^\top \mathbf{\mathfrak{N}} \ba]_k=\sum_{i=1}^{r} \sum_{j=1}^{r} \mathfrak{N}_{ijk} a_i a_j$. In Ref.~\cite{peherstorfer2016data}, this is rewritten as a matrix-vector product to exploit the commutative property of multiplication and avoid redundancy (i.e., we consider a single term $a_i a_j$ as a representative of both $a_ia_j$ and $a_j a_i$) as follows:
\begin{equation}
    \dfrac{\mathrm{d} \ba }{\mathrm{d}  t} = \mathbf{\mathfrak{B}} +  \mathbf{\mathfrak{L}} \ba +  \mathbf{\mathfrak{D}} \ba^2, \label{eq:opinf}
\end{equation}
where $\mathbf{\mathfrak{D}} \in \mathbb{R}^{r\times r(r+1)/2}$ is the quadratic operator and $\ba^2 = [\ba^{(1)^\top}, \ba^{(2)^\top}, \dots, \ba^{(r)^\top} ]^\top \in \mathbb{R}^{r(r+1)/2}$, with $\ba^{(i)} \in \mathbb{R}^i$ defined as
\begin{equation}
    \ba^{(i)} = a_i \begin{bmatrix} a_1 \\ \vdots \\ a_i \end{bmatrix}.
\end{equation}
Then, the components of $\mathbf{\mathfrak{B}}$, $\mathbf{\mathfrak{L}}$, and $\mathbf{\mathfrak{D}}$ are computed by solving $r$ least-squares problems, corresponding to each mode dynamics (i.e.,  $\dfrac{\mathrm{d} a_k }{\mathrm{d}  t}$). The OI algorithm in Ref.~\cite{peherstorfer2016data} can be extended to any arbitrary polynomial nonlinear terms in the state. However, the computational cost grows exponentially with the order of the polynomial nonlinear term rendering it feasible only for low-order polynomials. We highlight that even if the Galerkin projection step is not often required in the previous calibration (and OI) studies, it is generally assumed that the true governing equation has a quadratic (or polynomial) structure. This limitation was addressed in Ref.~\cite{qian2020lift}, which introduced a lift \& learn approach, where the ROM polynomial coefficients are efficiently calibrated even if the high-dimensional dynamics are not quadratic using lifting transformations. Recent OI developments are discussed in~\cite{benner2020operator,mcquarrie2021data,peherstorfer2020sampling,yildiz2021learning}.

\subsection{System identification approaches} \label{sec:id}

In many fluid dynamics applications, system identification approaches become viable tools to identify nonlinear low-order models \cite{loiseau2018constrained}. However, robust identification of realistic dynamical systems constitutes a grand challenge. For example, let us decompose the flow into two parts:
\begin{equation}
\mbox{Flow} = \mbox{resolved} + \mbox{unresolved dynamics}
\end{equation}
or
\begin{equation}
\boldsymbol{u} = \boldsymbol{v} + \boldsymbol{w}.
\end{equation}
Then, the evolution equation for $\boldsymbol{v}$ can be abstracted as follows:
\begin{equation}
\frac{d\boldsymbol{v}}{dt} = \boldsymbol f(\boldsymbol{v}, \boldsymbol{w}) 
\doteq \boldsymbol f(\boldsymbol{v}, 0) + \boldsymbol g(\boldsymbol{v}, \boldsymbol{w}). 
\end{equation}
The unresolved dynamics term, $\boldsymbol g$, has a high-frequency stochastic component important for short-term dynamics and an energy-absorbing component important for long-term boundedness. The first component can be modeled by a stochastic term, the second by an eddy viscosity model as we discussed earlier.

In general, data-driven model identification (without priors) requires full data. With full data a $k$-nearest neighbor ($k$-NN) model for kinematics and dynamics should work. With sparse data, no model identification might work. The lack of data has to be compensated by priors/knowledge. There are two issues here:

\emph{Warning 1.} Lack of resolution (unknown $\boldsymbol{w}$) leads to one closure problem. Lack of data (under-resolved $\boldsymbol{v}$) leads to another closure problem.

\emph{Warning 2.} The model complexity and data richness are strongly interwoven\citep{abu2012learning}. 

That being said, there are many methods for model identification, starting with brute-force data interpolation \cite{loiseau2018sparse}. Although many of the methodologies presented in Section~\ref{sec:da} can be viewed as system identification, we dedicate the following insights to the efforts and potential opportunities that aim at revealing the mathematical representation of the closure term. This is in contrast to assuming a specific form for model closure and fitting it to data to compute the unknown parameters and/or coefficients. In this regard, symbolic regression (SR) techniques have been recently exploited to identify interpretable closed form approximations of the governing equations, by observing the dynamical behavior and response of the system of interest \cite{quade2016prediction,luo2012parse}. SR techniques can be largely classified into two categories: (1) approaches that utilize compressed sensing and sparsity-promoting techniques to choose a few functions from a large feature library of potential basis functions that have the expressive power to define the dynamics; (2) evolutionary algorithms that search for functional abstractions with a preselected set of \emph{basic} mathematical operators and operands. Examples that belong to the first category include the sparse identification of nonlinear dynamics (SINDy) framework \cite{brunton2016discovering,de2020pysindy}, the sequential threshold ridge regression (STRidge) algorithm \cite{rudy2017data}, and the PDE-functional identification of nonlinear dynamics (PDE-FIND) technique \cite{rudy2017data}, while genetic programming (GP) \cite{koza1992genetic,bongard2007automated,schmidt2009distilling} and gene expression programming (GEP) \cite{ferreira2001gene} represent the major drivers for evolutionary SR explorations. 

\citet{loiseau2019pod} utilized SINDy to identify a system of nonlinearly coupled ODEs governing the evolution of the first pair of POD modes’ amplitudes (i.e., $a_1$ and $a_2$) for the 2D flow over a cylinder. Considering monomials of $a_1$ and $a_2$, a library of candidate functions is constructed as follows:
\begin{equation}
    \boldsymbol{\Theta} = \begin{bmatrix} 1 & a_1 & a_2 & a_1^2 & a_1 a_2 & a_2^2 & a_1^3 & a_1^2a_2 & a_1 a_2^2 & a_2^3 \end{bmatrix}.
\end{equation}
The identified ROM equations thus take the form
\begin{equation}
    \dfrac{\mathrm{d} \ba }{\mathrm{d}  t} = \boldsymbol{\Theta} \boldsymbol{\zeta} \label{eq:sindy},
\end{equation}
where $\boldsymbol{\zeta}$ encapsulates the coefficient of each candidate function, computed using a sparsity-promoting regression problem. Note that the library $\boldsymbol{\Theta}$ can be enriched with any arbitrary functions that potentially describe the system's dynamics. Since the system given in Eq.~\ref{eq:sindy} is distilled from data, the effect of truncated modes on $a_1$ and $a_2$ (i.e., the closure model) is inherited in the identified model. Indeed, the results of integrating the model derived by SINDy outperformed those from standard GROM.
A recent innovation\cite{Kaptanoglu2021arxiv},
trapping SINDy, identifies a dynamics with a trapping region, i.e., guarantees boundedness of the solution\cite{schlegel2015long}. 
This innovation is particularly important for higher-dimensional dynamics, 
where sparse identification is prone to give rise to unbounded solutions otherwise.

Symbolic regression approaches have been also pursued to discover \emph{discrepancy models} and reveal the hidden physics and dynamical processes that are not represented in the available governing equations. \citet{vaddireddy2020feature} applied the GEP and STRidge algorithms to recover the hidden physics (e.g., source or forcing terms) in the 2D Navier-Stokes equations using Eulerian sensor measurements. Likewise, \citet{kaheman2019learning} utilized SINDy to model the mismatch between simplified models and measurement data. With particular relevance to closure modeling, \citet{vaddireddy2020feature} also demonstrated the application of GEP and STRidge to identify the truncation error due to numerical discretization, and recover the eddy viscosity kernels, manifested as source terms in the LES equations. With the ongoing advancement and maturity of SR tools, great leaps in closure modeling are expected, especially given the lack of physical intuition in a ROM context.

\section{Physics-informed data-driven modeling}
\label{sec:pgml}

In most of the works related to NIROMs, the general idea is to employ the ML model to learn the temporal evolution of the field variable in the reduced order subspace and thereby bypass the intrusive Galerkin projection. \citet{gao2020non} proposed a framework that utilizes a fully connected neural network to approximate the nonlinear velocity function in the ROM equations by leveraging the same data used for the reduced basis construction. They illustrated the stable performance of the framework in the parametric viscous Burgers equation and two-dimensional premixed $H_2$-air flame model. \citet{xu2019multilevel} proposed a data-driven framework composed of a convolutional encoder to identify nonlinear basis functions, a temporal convolutional encoder to learn the temporal dynamics of latent variables, and a fully connected neural network to learn the mapping between the parameters of the system and the latent variables. They demonstrated the predictive performance of this framework for problems involving discontinuities, wave propagation, and coherent structures. Although NIROMs has been successful for many complex nonlinear problems, the typical sparsity of data motivates physics-informed data-driven modeling.     

\begin{remark}
We note that most computational modeling approaches require (a) data and (b) physics knowledge. The evolution equations are computationally too demanding and numerical discretization methods are not feasible for multi-query tasks. The data-driven methods are computationally efficient, but data-hungry in nature. For example, the autoencoders require a lot of data to build the generalizable ROM as it can not be derived from first principles. However, the reduced dynamics on the autoencoder need to follow first principle dynamics, which can be imposed as a prior given sparse data.
\end{remark}

The neural networks are one of the most popular algorithms for learning the reduced order dynamics of nonlinear problems. However, as the complexity of the system increases, the depth of the neural network also grows to learn the complicated nonlinear dynamics, and the number of trainable parameters explodes quickly. In the presence of sparse data, deep neural networks exhibit high epistemic uncertainty, and this adversely affects the trustworthiness of NIROMs. In these situations, it is important to inject physical relationships explicitly or implicitly in the training process or the model architecture. For example, partial differential equations, conservation laws, and symmetries can be exploited towards building physics-informed data-driven models. This will allow the training of deep neural networks with limited training data, speed up the training process, and also ensure physically consistent prediction. To this end, different approaches have been proposed in scientific machine learning, e.g., cost function modification to accommodate the model Jacobian \cite{lee1997hybrid}, grow-when-required network \cite{marsland2002self}, physics-informed neural networks \cite{raissi2019physics,zhu2019physics,pan2020physics,karniadakis2021physics}, embedding \emph{hard} physical constraints in a neural network \cite{mohan2020embedding,beucler2019achieving}, \textcolor{rev}{physics-based feature extraction \cite{meidani2021data}}, leveraging uncertainty information \cite{leibig2017leveraging}, developing visualization tool of the network analysis \cite{sacha2017you}, physics-guided machine learning \cite{pawar2021physics,pawar2021model}, and hybrid modeling\cite{pawar2020data,ahmed2020long,ahmed2020nudged,san2021hybrid}. \textcolor{rev}{Readers are referred to the recent review article by \citet{karniadakis2021physics} for a detailed discussion of embedding physics into machine learning for tackling scientific problems. In recent years, innovative applications of tailored neural network architectures have become increasingly common among fluid dynamicists. A central question in many studies is often how to exploit prior knowledge about the problem at hand to build more trustworthy models \cite{bonavita2021machine}. Various hybrid modeling principles that aim at combining machine learning and data-driven models with physics-based models have been recently discussed by \citet{san2021hybrid}.}

The physics-informed data-driven modeling has been successful in many applications, such as turbulence closure modeling \cite{ling2016reynolds}, super-resolution of turbulent flows \cite{subramaniam2020turbulence,BODE2021}, and generative modeling of dynamical systems \cite{erichson2019physics,wu2020enforcing,geneva2020modeling}, and it holds a great potential for ROMs. Next, we present several recent developments in physics-informed data-driven modeling. \citet{Chen2020PhysicsinformedML} built the physics-reinforced neural network (PRNN) trained by minimizing the mean squared residual error of the reduced order equations, and the mean squared error between the neural network prediction and the projection of high-fidelity data on the reduced basis. The incorporation of reduced order equations in the loss function can be considered as a physics-based regularization. The PRNN was demonstrated to be more accurate than a purely data-driven neural network for complex nonlinear problems. \citet{mohan2020embedding} proposed a physics-embedded convolutional autoencoder (PhyCAE) in which the divergence-free condition is imposed as a hard constraint through non-trainable layers after the decoder. The PhyCAE combined with a recurrent neural network for modal coefficients prediction can provide a physics-constrained NIROM that satisfies the conservation laws. \citet{lee2019deep} developed a framework that computes the lower-dimensional embedding using a convolutional autoencoder and enforces physical conservation laws by modeling the latent-dynamics as a solution to a constrained optimization problem. The objective function of the optimization problem is defined as the sum of squares of conservation-law violations over a control volume of the finite volume discretization. 

\textcolor{rev}{\citet{kaptanoglu2021physics} developed a physics-constrained low-dimensional model for magnetohydrodynamics by enforcing symmetries derived from global conservation laws into data-driven models. \citet{sawant2021physics} proposed a physics-informed regularizer and structure preserving (such as symmetry and definiteness in linear terms) formulation of OI and demonstrated their framework's performance in terms of improved stability and accuracy for nonlinear systems. The OI framework with physics-based regularization has also been applied for building predictive ROMs for rocket engine combustion dynamics \cite{swischuk2020learning}. A constrained sparse Galerkin approach has been introduced by \citet{loiseau2018constrained}. Finally, \citet{mohebujjaman2019physically} used physical constraints to increase the stability and accuracy of their data-driven variational multiscale ROM closure framework. }

\textcolor{rev}{There are a number of limitations of data-driven methods, including extrapolation beyond the training dataset, the curse of dimensionality, stability issues, and boundedness of the model. Many of these potential challenges can be mitigated using the decades of progress in physics-based modeling. For example, one way to address the curse of dimensionality, i.e., when the optimization problem can become highly non-convex and computationally expensive for very high-dimensional systems, is to tailor the feature space based on prior knowledge about the system \cite{beetham2020formulating}. Furthermore, the extrapolation beyond training dataset, which is a central challenge in many data-driven methods, can be effectively tackled by enforcing conservation laws or by using a custom neural network architecture that incorporates prior information about the system at hand. Finally, as the research in physics-informed data-driven modeling is rapidly progressing, reproducible research through open-source benchmark datasets and test cases should also be developed.}

\section{Conclusions and Outlook}

The ever-increasing need for computational efficiency and improved accuracy of many applications leads to very large-scale dynamical systems whose simulations and analyses make unmanageable demands on computational resources. Significantly simplifying the computational complexity of the underlying mathematical model, ROMs offer promise in many applications, like shape optimization, uncertainty quantification, and control. Over the past decades, for example, DNS, LES, and RANS have made a tremendous impact in the numerical simulation of turbulent flows. However, these FOMs cannot be generally used in such many query applications because of their prohibitively high computational cost. 

ROMs are efficient, relatively low-dimensional models often created from available data. In fluid dynamics, ROMs have been successfully used as surrogate models for structure-dominated problems, mainly in simple, academic test problems. However, traditional ROMs generally fail for more realistic flows because a low-dimensional ROM basis cannot accurately represent the complex dynamics, a topic that vastly waits for new explorations. Our paper provides a glimpse into various approaches to generating accurate ROMs, focusing on phenomenological, mathematical, and data-centric closure modeling approaches. We primarily focus on subspace projection-based methods and discuss their feasibility for various nonlinear problems in fluid dynamics. A chief emphasis is given to closure methods and, in particular, to the analogy between LES and ROMs. Various methodologies are leveraged and examples are included to provide a broader overview of forward-looking reduced order modeling practices in the age of data.  Of course, our paper is only a first step in an exhaustive discussion of ROM closure modeling.   Although we tried to include as many relevant contributions as possible, we left out important developments (e.g., stochastic closure modeling, compressible flows, and ROM pressure approximations).  We hope, however, that our paper will serve as a stepping stone toward a more comprehensive discussion of the exciting research areas of data-driven modeling and ROM closures, where these and many other new developments will be carefully presented.  



We envision that ROMs will be a key enabler in the development of a big data cybernetics infrastructure, an approach to controlling an asset or process using real time big data. These tools and concepts offer many new perspectives to our rapidly digitized society and its seamless interactions with many different fields. With the recent wave of digitalization, the latest trend in every industry is to build systems and approaches that will help it not only during the conceptualization, prototyping, testing, and design optimization phase but also during the operation phase with the ultimate aim to use them throughout the whole product life cycle. While the numerical simulation tools and lab-scale experiments are clearly important in the first phase, the potential of real time availability of data in the operational phase is opening up new avenues for monitoring and improving operations throughout the life cycle of a product. We believe that ROMs will be crucial to the improvement of emerging digital twin technologies.


Currently, the development of robust monolithic ROMs for a single operating condition with adequate data is a well-established art. We remark that `sufficient data' for post-transient dynamics implies that all snapshots will be approximately revisited multiple times. This is already quite a challenge for turbulent flows given the stochasticity of the dynamics.

A much more common task is modeling transient, controlled or multi-parametric dynamics. For instance, an airplane needs to be designed to prevent flutter under a large range of operating conditions, e.g., velocity, angle of ascent or descent, position of flaps, and  maneuvers. There will never be enough data for these cases and the large terra incognitae (oceans of missing data) have to be replaced by prior knowledge or clever guesses. Hence, physics-informed data-driven ROMs will become a necessity for most applications.

Most likely, such a range of dynamics will not be facilitated by a single `monolithic' reduced order representation, 
but a large set of representations, leading to a large set of ROMs with overlapping domains of validity. This trend is already foreshaddowed for the transient cylinder wake. An accurate reduced order representation requires about 50 POD modes \cite{loiseau2018sparse} or, staying in the Galerkin framework, a set of adjustable mean-field and adjustable Galerkin expansion modes \cite{Siegel2008jfm}.

Another requirement is human interpretability of the kinematics and dynamics. In an ideal case, this leads to a low-dimensional manifold for the data and a sparse representation for the dynamics. Most likely, only cartographic visualization of the state space and the dynamics will be achievable. One example is the  cluster-based network 
with many centroids for many operating conditions and a transition network for the dynamics.

As the treasures of high quality experimental and numerical data and the spectrum of increasingly powerful methods  are evolving, an open-source distribution of data and methods becomes of increasing importance. Hitherto, sharing of data and methods is still in its infancy and requires dedicated efforts. A noteworthy positive development is ever-increasing number of journals that encourage the publication of data and methods.

We conclude with two guidelines which are independent of the field but may easily be overlooked. First, there is the need for a well chosen set of guiding benchmark problems serving as lighthouses for model development. The second advice is best formulated by {Harrington Emerson} (1853--1931, efficiency engineer):
\begin{quote}
``As to methods there may be a million and then some, but principles are few. The man who grasps principles can successfully select his own methods. The man who tries methods, ignoring principles, is sure to have trouble.''
\end{quote}

\begin{acknowledgements}
O.S. gratefully acknowledges the U.S. DOE Early Career Research Program support DE-SC0019290 and the National Science Foundation support DMS-2012255.
A.R. acknowledges the financial support received by the Research Council of Norway and the industrial partners of the following projects: EXAIGON--{\em Explainable AI systems for gradual industry adoption\/} (grant no. 304843), {\em Hole cleaning monitoring in drilling with distributed sensors and hybrid methods\/} (grant no. 308823), and RaPiD--{\em Reciprocal Physics and Data-driven models\/} (grant no. 313909).
T.I. acknowledges support through the National Science Foundation grant DMS-2012253.
B.R.N. acknowledges funding from Harbin Institute of Technology, Shenzhen (Starting grant), the Peacock Talent A Plan from Shenzhen Government, 
and the German Science Foundation (DFG) via grant SE 2504/3-1.

\end{acknowledgements}


\section*{Data availability}
The data that supports the findings of this study are available within the article.

\bibliography{references}
\end{document}